\documentclass[3p]{elsarticle} % for review
\usepackage[T1]{fontenc}
\usepackage{lmodern}
\usepackage[latin1]{inputenc}
\usepackage{amsmath}
\usepackage{amsthm}
\usepackage{amsfonts}

\usepackage{multirow}
\makeatletter
\newskip\@bigflushglue \@bigflushglue = -3cm plus 1fil

\def\bigcentering{\let\\\@centercr\rightskip\@bigflushglue%
\leftskip\@bigflushglue
\parindent\z@\parfillskip\z@skip}

\makeatother
\usepackage{booktabs}
\newcommand{\egal}{\quad = \quad}
\newcommand{\ds}{\displaystyle}

\begin{document}

%% File Extensions of Graphics %%%%%%%%%%%%%%%%%%%%%%%%%%%%%%
%% ==> This enables you to omit the file extension of a graphic.
%% ==> "\includegraphics{title.eps}" becomes "\includegraphics{title}".
%% ==> If you create 2 graphics with same content (but different file types)
%% ==> "title.eps" and "title.pdf", only the file processable by
%% ==> your compiler will be used.
%% ==> pdfLaTeX uses "title.pdf". LaTeX uses "title.eps".
%\ifpdf
%	\DeclareGraphicsExtensions{.pdf,.jpg,.png}
%\else
%	\DeclareGraphicsExtensions{.eps}
%\fi
\graphicspath{{Figures/}}
\begin{frontmatter}


%% \title{}

\title{CN-Stream: Open-source library for nonlinear regular waves using stream function theory}


\author{Guillaume Ducrozet\corref{}}
\author{Benjamin Bouscasse\corref{}}
\author{Ma{\"i}t{\'e} Gouin\corref{}}
\author{Pierre Ferrant\corref{}}
\author{F{\'e}licien Bonnefoy\corref{cor1}}

\address{Ecole Centrale de Nantes, LHEEA Res. Dept. (ECN and CNRS) \\
1, rue de la No{\"e} - 44321 Nantes, France}

\cortext[cor1]{Corresponding author.\\E-mail address: felicien.bonnefoy@ec-nantes.fr\\Tel: +33 240 371 556\\Fax: +33 240 372 523}

\begin{abstract}
CN-Stream is a library for the computation of nonlinear regular ocean waves. The library is developed in order to be easily integrated with wave generation models in CFD solvers. It is based on the stream function theory and provides significant improvements regarding the applicability of the method for waves close to breaking (in deep or shallow water) compared to the classical implementation of Rienecker and Fenton \cite{Rienecker}. The complete description of the wave field is available, including the free-surface evolution and the wave kinematics in the fluid domain. It is released as open-source, developed and distributed under the terms of GPL v3. 
\newline

%HOS-ocean is an efficient High-Order Spectral code developed to solve the deterministic propagation of nonlinear wavefields in open ocean. HOS-ocean is released as open-source, developed and distributed under the terms of GNU General Public License (GPLv3). Along with the source code, a documentation under wiki format is available which makes easy the compilation and execution of the source files. The code has been shown to be accurate and efficient.
\end{abstract}


\end{frontmatter}

% \linenumbers

% Computer program descriptions should contain the following
% PROGRAM SUMMARY.

{\bf PROGRAM SUMMARY}
  %Delete as appropriate.

\begin{small}
\noindent
{\em Manuscript Title:} CN-Stream: Open-source library for nonlinear regular waves using stream function theory \\
{\em Authors:} G. Ducrozet, B. Bouscasse, M. Gouin, P. Ferrant, and  F. Bonnefoy \\
{\em Program Title:} CN-Stream                                         \\
{\em Journal Reference:}                                      \\
  %Leave blank, supplied by Elsevier.
{\em Catalogue identifier:}                                   \\
  %Leave blank, supplied by Elsevier.
{\em Licensing provisions:} GPL v3                                \\
  %enter "none" if CPC non-profit use license is sufficient.
{\em Programming language:} Fortran                                  \\
{\em Computer:} Tested on Intel Xeon E5504 and Intel Core i7 \\
  %Computer(s) for which program has been designed.
{\em Operating system:} Any system with a Fortran compiler: tested on Linux, OS X and Windows7 \\
  %Operating system(s) for which program has been designed.
{\em RAM:} Few MB for all configurations\\%From several MB up to several GB, depending on problem  (512x512, M=3: 385 MB) 256x256, M=3: 99MB   \\
  %RAM in bytes required to execute program with typical data.
%{\em Number of processors used:}                              \\
  %If more than one processor.
%{\em Supplementary material:}                                 \\
  % Fill in if necessary, otherwise leave out.
  % Please give some freely chosen keywords that we can use in a
  % cumulative keyword index.
%{\em Classification: 4.12 (and eventually: 7.7 \& 12)}                                         \\
  %Classify using CPC Program Library Subject Index, see (
  % http://cpc.cs.qub.ac.uk/subjectIndex/SUBJECT_index.html)
  %e.g. 4.4 Feynman diagrams, 5 Computer Algebra.
%{\em External routines/libraries:}     \\
  % Fill in if necessary, otherwise leave out.
%{\em Subprograms used:}                                       \\
  %Fill in if necessary, otherwise leave out.
%{\em Catalogue identifier of previous version:}*              \\
  %Only required for a New Version summary, otherwise leave out.
%{\em Journal reference of previous version:}*                  \\
  %Only required for a New Version summary, otherwise leave out.
%{\em Does the new version supersede the previous version?:}*   \\
  %Only required for a New Version summary, otherwise leave out.
{\em Nature of problem:}\\
  %Describe the nature of the problem here.
  CN-Stream has been developed to study the propagation of nonlinear regular waves over arbitrary constant water depth.
   \\
{\em Solution method:}\\
  %Describe the method solution here.
  CN-Stream is an implementation of the stream function method, which solves the problem by means of finite Fourier series to reduce the free-surface conditions to a set of nonlinear equations solved by Newton's iteration method. The algorithm provides an automatic choice of the optimal number of modal components for a given accuracy.
   \\
%{\em Reasons for the new version:}*\\
%  %Only required for a New Version summary, otherwise leave out.
%   \\
%%{\em Summary of revisions:}*\\
%  %Only required for a New Version summary, otherwise leave out.
%   \\
{\em Restrictions:}\\
  %Describe any restrictions on the complexity of the problem here.
  CN Stream is dedicated to the propagation of regular wave fields in infinite and finite constant depth. The evolution of irregular waves or over variable bathymetry is not treated. Furthermore, simulations are restricted to non-breaking waves.
   \\
%{\em Unusual features:}\\
%  %Describe any unusual features of the program/problem here.
%   \\
%{\em Additional comments:}\\
%  %Provide any additional comments here.
%   \\
{\em Running time:}\\
  %Give an indication of the typical running time here.
  {The solution is obtained in a running time of few seconds for all configurations.}
   \\
\end{small}

%\newpage
%\tableofcontents
%\newpage

\section*{Introduction}

The simulation of water waves is an old topic of investigation in naval and offshore hydrodynamics.  The knowledge of the incident wave field acting on a structure is important in the computation of loads. The linear description of waves is not sufficient for most realistic cases. An overview of some methods for the solution of regular waves in different conditions is presented in \cite{sobey1987application}. Regular waves are usually described either by the Stokes theory \cite{Stokes_1849} or the stream function theory, for instance the one presented by Rienecker and Fenton in \cite{Rienecker}.

\begin{figure}[!htbp]
	\centering
%		\includegraphics[scale=0.8]{Figures/Le-Mehaute-1976.png}
%	\caption{Le Mehaut\'e's diagram \cite{le_mehaute_introduction_1969}.}
		%\includegraphics[scale=1.0]{Figures/Le-Mehaute-1976.png}
		\includegraphics[scale=0.5]{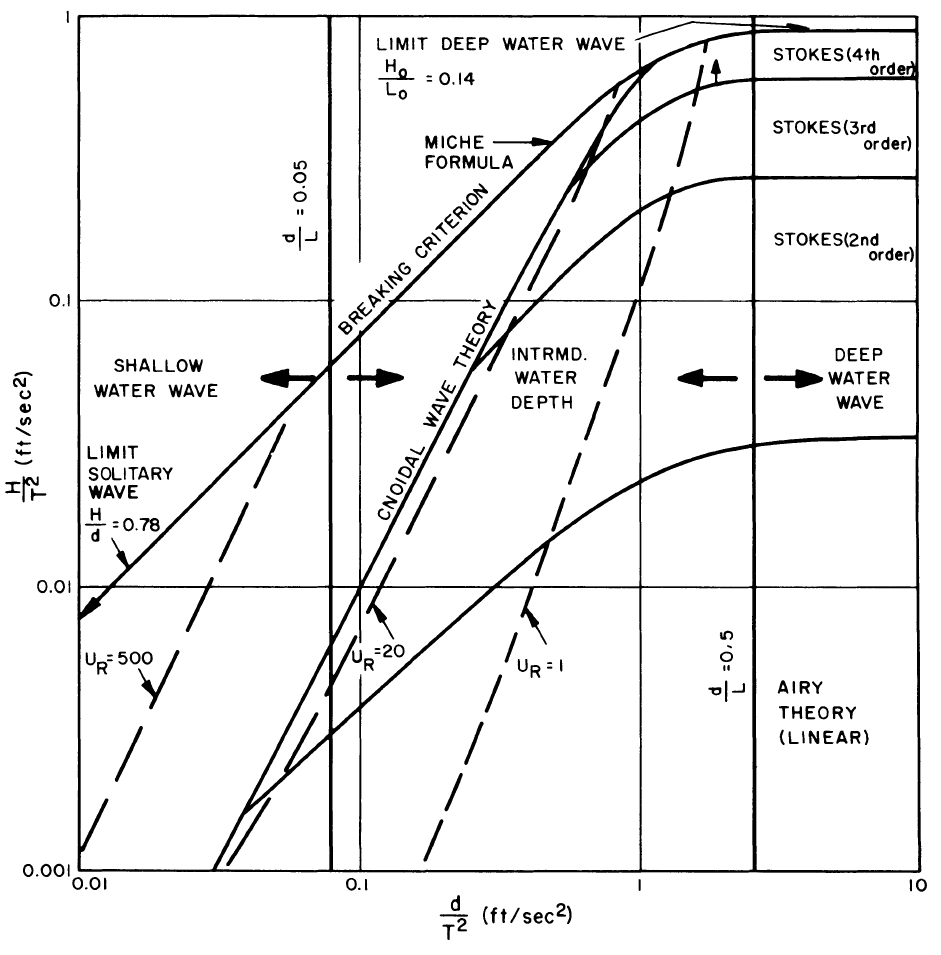}
	\caption{Le M\'ehaut\'e's diagram \cite{le_mehaute_1976}. $d$ stands for water depth, $L$ for wave length, $T$ for wave period and $U_R$ for Ursell number}
	\label{fig:Le_Mehaute}
\end{figure}

The important physical parameters in the context of wave propagation are the linear steepness $kH/2$ and the relative water depth $kh$, where $k=2\pi / \lambda$ is the wavenumber, $\lambda$ is the wavelength, $H$ is the wave height and $h$ is the water depth. A combination of these two parameters $H/h$ or the Ursell number $Ur=\frac{H\lambda^2}{h^3}=4\pi^2\frac{kH}{(kh)^3}$ can also be used, especially in the context of reduced water depth. In ocean engineering, the limits of applicability of the different wave theories are typically taken following Le Mehaut\'e's diagram \cite{le_mehaute_1976} presented in Fig.\ref{fig:Le_Mehaute}. In this figure, $d$ stands for water depth, $L$ for wave length, $T$ for wave period and $U_R$ for Ursell number.

%Bibliography review of stream function... Recent things existing? New method for limiting waves in arbitrary depth based on HAM \cite{zhong2017limiting}.

%"Up to now, there are only few methods (Schwartz 1974; Cokelet 1977; Williams 1981; VandenBroeck & Schwartz 1979) capable to solve the two-dimensional limiting (extreme) steady progressive wave in very shallow water. To the best of our knowledge, almost all analytic/numerical methods rely on the approximation techniques, such as the extrapolation and Pad�e approximant. Besides, almost all analytic/numerical methods fail to give accurate results for limiting waves in extremely shallow water depth, such as d/? < 0.005. Especially, to the best of our knowledge, accurate wave profiles for very shallow water are never presented previously, particularly in the case of d/? < 0.01. So, an analytic approach that can yield accurate results for the two-dimensional limiting (extreme) progressive gravity wave in arbitrary water depth without any kind of approximation techniques is of great meaning. This is the motivation of this paper." - zhong2017limiting

However, it is well established that the Stokes wave theory is not accurate for very steep waves or for shallow water depths. This perturbation method is not able to provide convergent high-order Fourier coefficients \cite{schwartz1974computer,cokelet1977steep}. The solution that is usually chosen is consequently to replace the perturbation expansions by a numerical evaluation, solving a nonlinear set of equations. This is assumed to be a more suitable approach for waves close to the wave breaking limit \cite{fenton1990nonlinear}. This enhanced accuracy is particularly important for the detailed physical analysis of such phenomena but also when looking for a reference solution for waves in nonlinear potential flow formalism. For instance, it is necessary to achieve such level of accuracy when propagating waves over a long time (\emph{e.g.} during 1000 waves periods as presented in \cite{Bonnefoy_ANSNWW_2010,Ducrozet_IJNMF_2011}) or when estimating the accuracy of a numerical model (see \cite{DUCROZET2016245}).

The original works \cite{chappelear1961direct,dean1965stream,chaplin1980developments,Rienecker} present different numerical solutions of the problem. The most widespread one in the ocean engineering community is probably the one described in details in \cite{Rienecker} and simplified in \cite{fenton1988numerical}. In \cite{fenton1988numerical}, the method is described and a Fortran program is provided.  It uses a finite Fourier series to reduce the free surface conditions to a set of nonlinear algebraic equations, and then used Newton's iteration method to solve these nonlinear equations. This is the one taken as basis in this work.
% Stream function theory --> Jacobsen et al (2012)

Note that other formulations and approaches exist with the main objective of increasing the accuracy for steep waves close to the wave breaking limit in arbitrary constant water depth. We can cite \emph{e.g.} \cite{vanden1979numerical} or the recent work of \cite{zhong2018limiting} which presents a numerical method free of any kind of approximation techniques or \cite{clamond2018accurate} that provides an efficient algorithm for computing steady surface gravity waves for all wavelength over depth ratios.
\newline

%"VandenBroeck & Schwartz (1979) proposed an efficient numerical scheme to solve the steep gravity wave. They first formulated the steep gravity waves as a system of integrodifferential equations, and then used the Newton?s iteration technique to solve the coupled equations. Using this numerical method, accurate results can be obtained even in the case of d/? = 0.008."

%Possibly make a short review of the use of stream function in nonlinear potential flow solvers, CFD, etc.

CN-Stream is an open-source stream function model developed at Ecole Centrale Nantes, LHEEA Res. Dept. (ECN and CNRS). The software is available to download and contribute on the GitHub platform \cite{CN-Stream}. The code is developed and redistributed under the terms of the licence GPL v3. Documentation that describes the compilation and execution of the source files is provided along with the source code. This code is one of the open-source wave models developed at Centrale Nantes. Others wave generation codes available on the GitHub platform are HOS-ocean \cite{DUCROZET2016245} , HOS-NWT \cite{ducrozet2012modified} respectively for 2D and 3D non linear wave generation in open water and wave basin and Grid2Grid \cite{choi2017grid2grid}, which serves for their coupling with CFD (Computational Fluid Dynamics) methods.
\\

In the following sections, the stream function theory and the corresponding numerical procedure are briefly presented together with the improvements proposed and implemented in CN-Stream. Sections describing how to compile and use the code as a library are also provided. Finally, different study cases are presented as typical applications of the presented numerical model.

One of the purposes of this code is to encourage other researchers to use this library in the context of the coupling with CFD software for wave-structure interactions modeling. Indeed, in a lot of nonlinear potential flow solvers \cite{DUCROZET2016245,engsig-karup_efficient_2009} or CFD softwares (SPH \cite{oger_coupled_2014}, WCCH \cite{li_VOF_2017}, ICARE \cite{reliquet_simulation_2013}, OpenFOAM \cite{jacobsen2012wave}), the incident waves are issued of the stream function theory. Some examples of the reconstructed volume fields used in CFD models are provided.

\section{Stream function method}\label{sec:Stream_Function}

In this section, the formulation of the problem to obtain the nonlinear solution is presented in a simplified manner. More details can be found in the original work of \cite{Rienecker} or \cite{fenton1988numerical}, taken as basis for the numerical model CN-Stream. Some improvements of the original numerical method are then detailed.

\subsection{Coordinate system}

The wave propagation is solved in a fixed reference frame $(O,{X},{Z})$ with the origin O taken on the free surface at rest: the horizontal axis $X$ is oriented in the direction of the waves, and the $Z$ axis is vertical upward.

The wave solution of the problem is assumed to be periodic both in space and time. The free surface profile is of permanent shape and the wave is propagating with a constant phase velocity $c$. The solution becomes stationary in a moving reference frame denoted as $({x},{z})$. The horizontal axis $x$ is oriented in the direction of wave propagation and the vertical axis $z$ is upward with the origin at the free surface at rest.
\begin{eqnarray}
x & = &X-ct \\
z & = &Z 
\end{eqnarray} 

Note that in the original article of \cite{Rienecker}, the origin of the vertical axis was located on the sea bed. This induces the following changes with respect to this initial work:
\begin{eqnarray}
y & \longrightarrow & z+h \\
R & \longrightarrow & R+h \\
Q & \longrightarrow & Q+hb_0 
\end{eqnarray}

The exact definition of the different variables ($R, Q, b_0$) is presented in the following sections. %\\

%\textbf{Is it a problem to introduce those changes before the definition of variables?}

%The origin of the coordinate system is located at the free surface at rest: this is the convention used here. In the original article of \cite{Rienecker}, the origin was located on the sea bed. The following changes have been made with respect to this initial work:
%\begin{eqnarray}
%y & \longrightarrow & z+h \\
%R & \longrightarrow & R+h \\
%Q & \longrightarrow & Q+hb_0 
%\end{eqnarray} 
%It is moving at the phase velocity $c$ with respect to a fixed frame. The horizontal axis $x$ is oriented in the direction of wave propagation and the vertical axis $z$ is upward.
%In the case of CN-stream, we express the solution in a fixed reference frame which coordinates are denoted $(X,Z)$. Thus the following changes are made:
%\begin{eqnarray}
%x =&X-ct \\
%z=&Z \\
%B_0 =& c+b_0 
%\end{eqnarray} 
%%We simply have to replace the horizontal space variable $x$ by $X-ct$ and the linear part (in $x$) of the potential $b_0$ by $c+b_0$. 
%The origin of the fixed reference frame is taken on the free surface at rest, the horizontal axis $X$ is oriented in the direction of the waves, and the $Z$ axis is vertical upward.
%
%
%

\subsection{Equations}
In the case of bi-dimensional isovolume flow, the stream function $\psi(x,z)$ allows for the representation of the velocity field $\textbf{V}=(u,w) = (\frac{\partial \psi}{\partial z} ,-\frac{\partial \psi}{\partial x} ) $.

Furthermore, if the motion is irrotational, $\psi$ satisfies the Laplace's equation in the fluid domain:
\begin{equation}
\Delta \psi =0
\label{eq:Laplace}
\end{equation}

The free surface elevation is defined as $z=\eta(x)$ and the different boundary conditions are:
\begin{itemize}
	\item the dynamic free surface boundary condition:
   \begin{equation}
   g\eta + \frac{1}{2} \, \left[ \left(\frac{\partial \psi}{\partial x}\right)^2 + \left(\frac{\partial \psi}{\partial z}\right)^2 \right] \egal R\ \text{,~~~~~ on } z=\eta
   \label{eq:CSLD}
   \end{equation}
   with $R$ the so-called Bernoulli constant,
   \item a free-slip condition on the free surface $z=\eta(x)$ (also known as kinematic free surface boundary condition)
   \item and a free-slip condition on the bottom $z=-h$.
\end{itemize}

The free-slip boundary conditions are easily written with the stream function, which has the following properties:
\begin{itemize}
	\item iso-lines represent the streamlines,
	\item the variation of the stream function between two streamlines is equal to the flow rate between those lines.
\end{itemize}

The bottom of the domain and the free surface being streamlines when considering the moving reference frame at phase velocity (and consequently permanent elevation), it is chosen to impose at the bottom:
\begin{equation}
\psi(x,z=-h) \egal 0.
\label{eq:BBC}
\end{equation}
As a consequence, the stream function at the free surface is related to the flow rate $Q$ between the bottom and the free surface. This gives:
\begin{equation}
\psi(x,z=\eta(x)) \egal - Q.
\label{eq:psiQ}
\end{equation}
In addition, the free surface presents a zero mean elevation with respect to the definition of the origin of the vertical axis. This is written as:
\begin{equation}
\int_0^\lambda \eta(x) dx \egal 0.
\end{equation}

\noindent Then, $\eta$ and $\psi$ can be decomposed with the help of Fourier series in the horizontal plane:
\begin{equation}
\eta(x) \egal \frac{a_0}{2} + \sum_{n=1}^{+\infty} \, a_n \, \cos (k_nx)
\label{eq:etaRFinit}
\end{equation}
\begin{equation}
\psi(x,z) \egal b_0z + \sum_{n=1}^{+\infty} \, b_n \, \frac{\sinh (k_n(z+h))}{\cosh (k_nh)} \, \cos (k_nx)
\label{eq:psiRF}
\end{equation}
with $a_n$ and $b_n$ the modal amplitudes of the free surface elevation and the stream function respectively. In the moving reference frame $(x,z)$, those are constant for a given wave. This equation satisfies both Eq.\eqref{eq:Laplace} and Eq.\eqref{eq:BBC}.
\newline

Equivalently, we can write the horizontal velocity $u$ and the vertical velocity $w$:
\begin{equation}
u(x,z) \egal b_0 + \sum_{n=1}^{+\infty} \, k_n b_n \, \frac{\cosh (k_n(z+h))}{\cosh (k_nh)} \, \cos (k_nx)
\end{equation}
\begin{equation}
w(x,z) \egal  \sum_{n=1}^{+\infty} \, k_n b_n \, \frac{\sinh (k_n(z+h))}{\cosh (k_nh)} \, \sin (k_nx)
\end{equation}
and the pressure is defined as:
\begin{equation}
\frac{p(x,z)}{\rho} \egal R - gz -\frac{1}{2}\left[u^2(x,z) + w^2(x,z)\right]
\end{equation}

\noindent
with $\rho$ the water density.
 %Eq.\eqref{eq:CSLD} and  Eq.\eqref{eq:psiQ} can then be written by replacing 
%
%
%
\subsection{Numerical solution}
\subsubsection{Inputs}
The numerical solution needs some inputs that will define the wave to be solved. Different choices are possible for its description and the corresponding inputs are:
\begin{itemize}
	\item the wave length $\lambda$ or the wave period $T$
	\item the wave height $H$
	\item the water depth $h$ (finite or infinite)
	\item the value of the current $U_c$ that may be of two kinds: i) a Eulerian transport (the reference frame moves with respect to the fixed reference frame) or ii) a fixed mass transport velocity.
\end{itemize}
The inputs can be in dimensional or non-dimensional form.

\paragraph{Dimensional wave parameters}
In the case of dimensional inputs, the required parameters are given in Tab.\ref{tab:dim} depending on the water depth and the known wave parameter ($T$ or $\lambda$).
\begin{table}[h!tbp]
   \centering
   \begin{tabular}{@{}lcc@{}}
      \toprule
      Known parameter & Infinite depth & Finite depth $h$ \\
      \midrule
      Period $T$ & $\ds (T,H,U_c)$ & $\ds (T,H,h,U_c)$ \\
      \addlinespace
      Wave length $\lambda$ & $(\lambda,H,U_c)$ & $(\lambda,H,h,U_c)$ \\
      \bottomrule
   \end{tabular}
   \caption{Dimensional input parameters for CN-Stream.}
	\label{tab:dim}
\end{table}

\paragraph{Non-dimensional wave parameters}

In the case of non-dimensional value, we set non-dimensional wave height, water depth and current, denoted respectively $H'$, $h'$ and $U'_c$. They are defined as follows (Tab. \ref{tab:adim}), depending if we know/fix as input the period or the wavelength. Note that the linear theory gives the simple following relations between the two sets of non-dimensional parameters for the wave height and the water depth:
\begin{equation}
\begin{array}{c}
\displaystyle H' \egal \frac{H}{gT^2} \egal  \, \frac{k_L H}{4\pi^2} \\
\displaystyle h' \egal  \frac{h}{gT^2} \egal \,  \frac{ k_L h}{4\pi^2}
\end{array}
\end{equation}
where $k_L$ indicates the wave number obtained from linear dispersion relation, which is consequently slightly different from the exact wave number (see Sec. \ref{subsec:Nonlinear_effects}). However, for an estimate, we can also set $kH \simeq 40 H'$ and $kh \simeq 40h'$.

\begin{table}[h!tbp]
   \centering
   \begin{tabular}{@{}lcc@{}}
      \toprule
      Known parameter & Infinite depth & Finite depth $h$ \\
      \midrule
      Period $T$ & $\ds \left(H'=\frac{H}{gT^2},U'_c=\frac{U}{gT}\right)$ & $\ds \left(H'=\frac{H}{gT^2},\ds h'=\frac{h}{gT^2},U'c=\frac{U}{gT}\right)$ \\
      \addlinespace
      Wave length $\lambda$ & $\ds \left(H'=kH,U'_c=\frac{U}{\sqrt{g/k}}\right)$ & $\ds\left(H'=kH,h'=kh,U'_c=\frac{U}{\sqrt{g/k}}\right)$ \\
      \bottomrule
   \end{tabular}
   \caption{Non-dimensional input parameters for CN-Stream.}
	\label{tab:adim}
\end{table}

\subsubsection{Non-dimensional wave outputs}

In the case of non-dimensional input values as described in previous section, the outputs are also made non-dimensional. This is dependent on the input:
\begin{itemize}
	\item Wavelength as input: length scale $L_s=\lambda$ and time scale $T_s=\sqrt{\lambda/g}$
	\item Period as input: length scale $L_s=gT^2$ and time scale $T_s=T$
\end{itemize}

All lengths are consequently non-dimensional with length scale, for instance $\eta'=\eta/L_s$ ; velocities with $U'=UT_s/L_s$ ; etc.

\subsubsection{Discretization}
The free surface elevation can be studied considering its $N_2+1$ values at the collocation points or
equivalently by expressing it on $N_2+1$ modes of the Fourier series:
\begin{equation}
\eta(x) \egal \frac{a_0}{2} + \sum_{n=1}^{N_2} \, a_n \,  \cos (k_nx)
\end{equation}

For the stream function, its representation in Fourier series is truncated at another number of modes chosen as $N_1+1$:
\begin{equation}
\psi(x,z) \egal b_0z + \sum_{n=1}^{N_1} \, b_n \, \frac{\sinh (k_n(z+h))}{\cosh (k_nh)} \, \cos (k_nx)
\end{equation}

The independent choice of $N_1$ and $N_2$ is one of the main difference with the original algorithm \cite{Fenton_1988}. The motivation and implications of this choice will be detailed in Sec. \ref{Results}.\\

Collocation points $x_m$ are defined with respect to the free surface, fixed to a number $N_2+1$ between the crest and the trough of the wave (a vertical symmetry exists on half a wavelength). Previous set of equations is discretized on those collocation points such as $z=\eta(x_m)$ with $x_m=m\frac{\lambda}{2 N_2}$ ($m\in[0,N_2]$).
\subsubsection{Unknowns}
%
%
%

%In all configurations, the unknowns are the modal amplitudes of the stream function and the free surface elevation namely $b_n$ for $n=0$ to $N_1$ and $a_n$ for $n=0$ to $N_2$ respectively, the constants $R$ and $Q$ and the phase velocity $c$.\\
In all configurations, the unknowns are the modal amplitudes of the stream function $b_n$ for $n=0$ to $N_1$  and the free surface elevation $\eta(x_m)$ for $m=0$ to $N_2$, the constants $R$ and $Q$ and the phase velocity $c$.\\

In addition we have:
\begin{itemize}
	\item The wave number $k$ if we specify as input the wave period $T$,
	\item The wave period $T$ if we specify as input the wave number $k$,
%    \item The wave height if we give as input the steepness $kH/2$ or $\varepsilon=H/\lambda$.
\end{itemize}

This corresponds to a total number of unknowns of $N_1+N_2+6$.\\

%\vspace{0.5em}
\subsubsection{Equations}
These unknowns satisfy, at a given accuracy, the following discrete nonlinear equations:
\begin{itemize}
   \item the dynamic free surface boundary condition (Eq. \eqref{eq:CSLD}) written at the collocation points,
    \begin{equation} 
     g \eta(x_m) + \frac{1}{2} \, \left[ u(x_m,\eta(x_m))^2 + w(x_m, \eta(x_m))^2 \right] \egal R\  \,,  \; \; m\in[0,N_2]
   \end{equation}
   
   \item the kinematic free surface boundary condition (Eq. \eqref{eq:psiQ}) written at the collocation points,
   \begin{equation} 
   \psi(x_m,\eta(x_m)) \egal - Q \,,  \; \; m\in[0,N_2]
   \end{equation}
   
   \item the zero-mean free surface elevation, which is written using trapezoidal rule:
\begin{equation}
0 \egal \eta(x_{0}) + \eta(x_{N_2}) + \sum_{m=1}^{N_2-1} \eta(x_{m}),
\label{eq:zero_mean}
\end{equation}
   \item the fixed wave height 
\begin{equation} 
   H=\max (\eta) - \min (\eta) = \eta(x_{0}) - \eta(x_{N_2})
   \label{eq:wave_height}
\end{equation}
   %\item the null Eulerian velocity $c+b_0=0$.
\end{itemize}

The stream function is built so that $b_0$ is the mean velocity of the fluid in the reference frame linked to the wave, moving at the phase velocity $c$. The method allows to take into account the influence of a current of two kinds:
\begin{itemize}
   \item Eulerian transport (the reference frame is moving at a velocity $c_E$ with respect to the fixed reference frame). This leads to the following equation: $c_E=c+b_0$,
   \item mass transport: $c_S=c-Q/h$.
\end{itemize}

\noindent One last equations is needed to close the system, which uses the relationship between $k$, $c$, and $T$:
%\begin{itemize}
%   \item the relationship between $k$, $c$, and $T$:
   \begin{equation}
    k \, c \, T \, =    \, 2\pi,
    \label{eq:relationkcT}
   \end{equation}
%   \item the relationship between $H$ and $k$ (or $H$ and $\lambda$)  if we give as input the steepness (in the form of $kH/2$ or $\epsilon$):
%   \begin{equation}
%   H=2 \, kH/2/k \ \text{or} \ H=\varepsilon \lambda.
%   \end{equation}
%\end{itemize}

We consequently end up with $2N_2+6$ equations.

%Au LHEEA, on se place dans le rep�re fixe \ie\ on impose simplement $c_E=0$.

\subsubsection{Numerical scheme}
\label{subsubsec:Numerical_scheme}

%In a nutshell, if we specify the wave length, the unknowns are the modal amplitudes $a_n$ for $n=0$ to $N_1$ and $b_n$ for $n=0$ to $N_2$, the phase velocity $c$, the variables $R$ and $Q$ and the period $T$, corresponding to a total number of unknowns of $N_1+N_2+6$.

%The equations \eqref{eq:CSLD} and \eqref{eq:psiQ} written at the collocation points give $2N_2+2$ equations. Assuming $N_2 \geq N_1$, the solution should work for $N_1=N_2$ (this is the original formulation of \cite{Rienecker}). We consequently need in addition the equations for the wave height (Eq. \eqref{eq:wave_height}), the volume (Eq. \eqref{eq:zero_mean}), the equation for current definition and the equation \eqref{eq:relationkcT} for the wave number or the period.\\

We assume for the numerical solution of the problem that $N_2 \geq N_1$. The system is consequently over-defined with $2N_2+6$ equations and $N_1+N_2+6$ unknowns.  Initial values for $Q$ and $R$ have to be given to solve the problem.

In \cite{Rienecker}, the particular case $N_1=N_2$  is solved iteratively  with a Newton-Raphson method, while least square method is used in the present implementation. The different equations to solve are expressed under the form $f(\eta(x_m),b_n,c,R,Q,T \mbox{ or }k) = 0$.  The system is linearized at each iteration $i$  to obtain an equation of the form: 
\begin{equation}
\label{iterativeMatrix}
A \left(Z^{i+1} - Z^i \right)=F^i
\end{equation}
where $A$ is the Jacobian matrix formed with the derivative of the equations with respect to the different variables, $Z^i$ the solution vector and $F^i$ an error vector.

If absolute errors are retained, the convergence of the solution is determined with thresholds set on $\epsilon^{abs}_Z = Z^{i+1} - Z^i$ and  $\epsilon^{abs}_F = F^i - F^{i-1}$.
If the convergence is controlled with relative errors, those are defined as $\epsilon^{rel}_Z = \dfrac{Z^{i+1} - Z^i}{\text{Scale}(Z^i)}$ and $\epsilon^{rel}_F = \dfrac{F^i - F^{i-1}}{\text{Scale}(F^{i-1})}$, where the function ``Scale'' ensure that the first modes are the one giving the magnitude of the solution.

%It is usually preferred to specify as inputs:
%\begin{itemize}
%	\item the wave height $H$ (or its steepness $ka$ or $\varepsilon=\pi H/\lambda$),
%	\item the wave length $\lambda$ or the wave period $T$ with the relation $\lambda=cT$.
%\end{itemize}
%
%
%
\subsubsection{Initial solution}
The first order Stokes solution was used in \cite{Rienecker} as the initial solution. Here we choose to impose the second-order Stokes solution, which gives the free surface elevation as:
\begin{equation}
\eta(x) \egal \frac{H}{2} \, \cos kx + k\left(\frac{H}{2}\right)^2 \, \frac{3 - \sigma^2}{4\sigma^3}\, \cos 2kx, 
\end{equation}
with $\sigma=\tanh kh$. The stream function at the free surface is defined as:
\begin{equation}
\psi(x,z=\eta) \egal -c + \frac{gH}{2kc} \, \sin kx + 3k\left(\frac{H}{2}\right)^2 \, \frac{1 - \sigma^2}{4\sigma^3}\, \sin 2kx. 
\end{equation}
$Q$ is set to $Q=0$ and $R$ to $R=-c^2/2$.

\subsection{From stream function to velocity potential}
From the definition of the velocity potential and the stream function, we have the following equalities:
\begin{equation}
\frac{\partial \phi}{\partial x}\egal \frac{\partial \psi}{\partial z},
\end{equation}
and
\begin{equation}
\frac{\partial \phi}{\partial z}\egal -\frac{\partial \psi}{\partial x}.
\end{equation}

The velocity potential $\phi$ is thus defined in the moving reference frame $(x,z)$ as:
\begin{equation}
\phi(x,z) \egal b_0x + \sum_{n=1}^{N_1} \, b_n \, \frac{\cosh (k_n(z+h))}{\cosh (k_nh)} \, \sin (k_nx) \ \text{.}
\label{eq:phi}
\end{equation}

When going back to the fixed reference frame (O,{X},{Z}), the problem becomes non-stationary. We remind that capital letters refer to the fixed reference frame, while small letters refer to the moving one, with the following change of coordinates:
\begin{eqnarray}
x &=&X-ct \\
z &=&Z \\
b_0 &=& c+B_0
\end{eqnarray}

\noindent In the fixed grid the elevation $\eta$ and the velocity potential $\phi$ are thus defined as:
\begin{equation}
\eta(X,Z,t) \egal \frac{a_0}{2}+\sum_{n=1}^{N_2} \, a_n \, \cos (k_n(X-ct))
\label{eq:etaRF}
\end{equation}
\begin{equation}
\phi(X,Z,t) \egal (c+B_0)(X-ct) + \sum_{n=1}^{N_1} \, b_n \, \frac{\cosh (k_n(Z+h))}{\cosh (k_nh)} \, \sin (k_n(X-ct))
\end{equation}
And the slope used in section \ref{slope} is simply defined as:
\begin{equation}
\frac{\partial \eta}{\partial X}(X,Z,t) \egal - \sum_{n=1}^{N_2} \, a_n \, k_n \sin (k_n(X-ct))
\label{eq:detadxRF}
\end{equation}

The horizontal velocity U and vertical velocity W are thus written as:
\begin{eqnarray}
U(X,Z,t) &&\egal c + B_0 + \sum_{n=1}^{N_1} \, k_n b_n \, \frac{\cosh (k_n(Z+h))}{\cosh (k_nh)} \, \cos(k_n(X-ct)) \\
&& \egal c + u(X-ct, Z)
\end{eqnarray}
\begin{eqnarray}
W(X,Z,t)&& \egal  \sum_{n=1}^{N_1} \, k_n b_n \, \frac{\sinh (k_n(Z+h))}{\cosh (k_nh)} \, \sin (k_n(X-ct))\\
&& \egal w(X-ct, Z)
\end{eqnarray}
and the pressure P:
\begin{eqnarray}
\frac{P(X,Z,t)}{\rho} &&\egal R - gZ -\frac{1}{2}\left[ u^2(X-ct, Z)+w^2(X-ct, Z)\right] \\
&&\egal  R - gZ -\frac{1}{2}\left[ (U(X,Z,t)-c)^2 +W^2(X, Z,t)\right]
\end{eqnarray}
Using:
\begin{equation}
\frac{\partial \phi}{\partial t}(X,Z,t) = -c \frac{\partial \phi}{\partial X}(X,Z,t)=-c U(X,Z,t)
\end{equation}
it comes:
\begin{equation}
\frac{P(X,Z,t)}{\rho} \egal R - gZ - \frac{1}{2}c^2-\frac{\partial \phi}{\partial t}(X,Z,t)  -\frac{1}{2}\left[U^2(X,Z,t) + W^2(X,Z,t)\right]
\end{equation}

\subsubsection{Remarks}
For some applications (see \emph{e.g.} \cite{DUCROZET2016245}), the dynamic free surface boundary condition Eq.\eqref{eq:CSLD} is written in terms of the velocity potential  $\tilde{\phi}(X,Z,t))$ under the following form:
   \begin{equation}
   \frac{\partial \tilde{\phi}}{\partial t}- g\eta - \frac{1}{2} \, \left[ \left(\frac{\partial \tilde{\phi}}{\partial X}\right)^2 + \left(\frac{\partial \tilde{\phi}}{\partial Z}\right)^2 \right] \egal 0\ \text{,~~~~~~ on } Z=\eta
   \label{eq:CSLD_phi}
   \end{equation}

This equation differs from Eq.\eqref{eq:CSLD} in terms of the gauge condition imposed to uniquely define the velocity potential, see \cite{clamond2017remarks}. The velocity potential $\phi(X,Z,t)$ does not satisfy the new dynamic boundary condition Eq.\eqref{eq:CSLD_phi}, leading to the definition of another velocity potential, namely $\tilde{\phi}$. The latter has to satisfy the following equation
\begin{equation}
 \frac{\partial \tilde{\phi}}{\partial t}(X,Z,t)= \frac{\partial \phi}{\partial t}(X,Z,t)-R +\frac{1}{2}c^2,
\end{equation}
leading to:
\begin{equation}
\tilde{\phi}(X,Z,t)= \left(-R +\frac{1}{2}c^2\right)t +(c+B_0)X +\sum_{n=1}^{N_1} \, b_n \, \frac{\cosh (k_n(Z+h))}{\cosh (k_nh)} \, \sin (k_n(X-ct))
\end{equation}
Note that to keep the spatial periodicity of the potential in the $x$-direction, the following condition needs to be satisfied:
\begin{equation}
c+B_0=0.
\end{equation}
This condition is satisfied if the Eulerian velocity $c_E$ is taken equal to zero.
\subsection{Improvements}
\label{improvements}
%The main improvements regarding the original solution of \cite{fenton1988numerical} concerns the optimal choice of the number of modes. Indeed, differently from the solution of \cite{Rienecker}, the number of modes for the description of $\eta$ and $\psi$ are not identical in CN-Stream. The choice of the number of modes $N_2+1$ and $N_1+1$ is of major importance for the numerical solution of the free surface boundary conditions as demonstrated in the results section \ref{Results}. Enhanced convergence properties are achieved compared to the original implementation.
%
%Moreover, an automatic calculation of the optimal number of collocation points (or equivalently of the number of modes) has been implemented. It consists in adapting the value of $N_1$ (number of modes for the stream function (or eq. velocity potential)) automatically so that the amplitude of the last mode reaches the accuracy given in input by the user. Then, the number of modes $N_2$ of the elevation is deduced from $N_1$ by an empirical formula derived later in this paper (see section \ref{N1_N2}).

\subsubsection{Increments in wave height}

When considering waves very close to the wave breaking limit, it appears that the numerical procedure may have some difficulty to converge toward a proper solution. In order to overcome this issue, the solution is looked for as an iterative process on the target wave height $H$.

The idea is to increase gradually the height of the non-linear wave, toward the final target one. At the end of one iteration, the non-linear solution for a given wave height is taken as the intial solution for the next iteration (\emph{i.e.} a higher wave height). This allows to find an accurate solution for non-linear waves very close to the wave breaking limit, as detailed in Sec. \ref{subsubsec:limiting_waves}.

The necessity of such procedure is actually related to the fact that: i) the choice of the number of modes $N_1$ and $N_2$ should be adequate to the simulated wave and ii) the second order solution is not accurate enough for highly non-linear wave. The solution procedure needs an initial guess close enough to the fully non-linear solution to be convergent.\\

Then, the user can specify as input the number of steps in the wave height (variable $n_H$ of the input file), together with an increment type for these wave heights, which is either linear or exponential. As a summary, the different successive wave heights are defined as follows with $H_t$ the target wave height, $n_H+1$ the number of steps and $i_H \in [1, n_H+1]$ the index of the iteration:
\begin{itemize}
	\item Linear increment:
	\begin{equation}
	H(i_H) = \frac{i_H}{n_H+1} H_t
	\end{equation}
	\item Exponential increment:
	\begin{equation}
	H(i_H) = 0.01 H_t + 0.99 H_t \log \left( 1 + \frac{i_H-1}{n_H} \left[e^1 - 1\right] \right)
	\end{equation}
\end{itemize} 

\subsubsection{Automatic evaluation of $N_1$ and $N_2$}
\label{subsec:automatic}

When solving the problem, an automatic evaluation of the optimal number of collocation points (or equivalently of the number of modes) is performed. Together with the independent choice of the number of modes for the descritpion of the stream function (or eq. velocity potential) $N_1$ and free-surface elevation $N_2$, these represent the main enhancements of the present numerical solution compared to the original one of \cite{Rienecker}.

This routine is called after the solution of the linear system (achieved with a least square method) which uses specific numbers of modes $N_1$ and $N_2$ (see Sec. \ref{subsubsec:Numerical_scheme}). Then, the number of modes is adjusted with the procedure described hereafter, leading to a new linear system (solved as in the previous step) until the convergence criteria on the choice of the number of modes is reached.
\\

The algorithm consists in adapting the value of the number of modes for the description of the stream function (or eq. velocity potential) $N_1$ automatically so that the amplitude of the last mode is smaller than the target accuracy provided by the user, denoted $\epsilon_{N_1}$. 

\begin{figure}[h!tbp]
	\centering
		\includegraphics[width=0.7\textwidth]{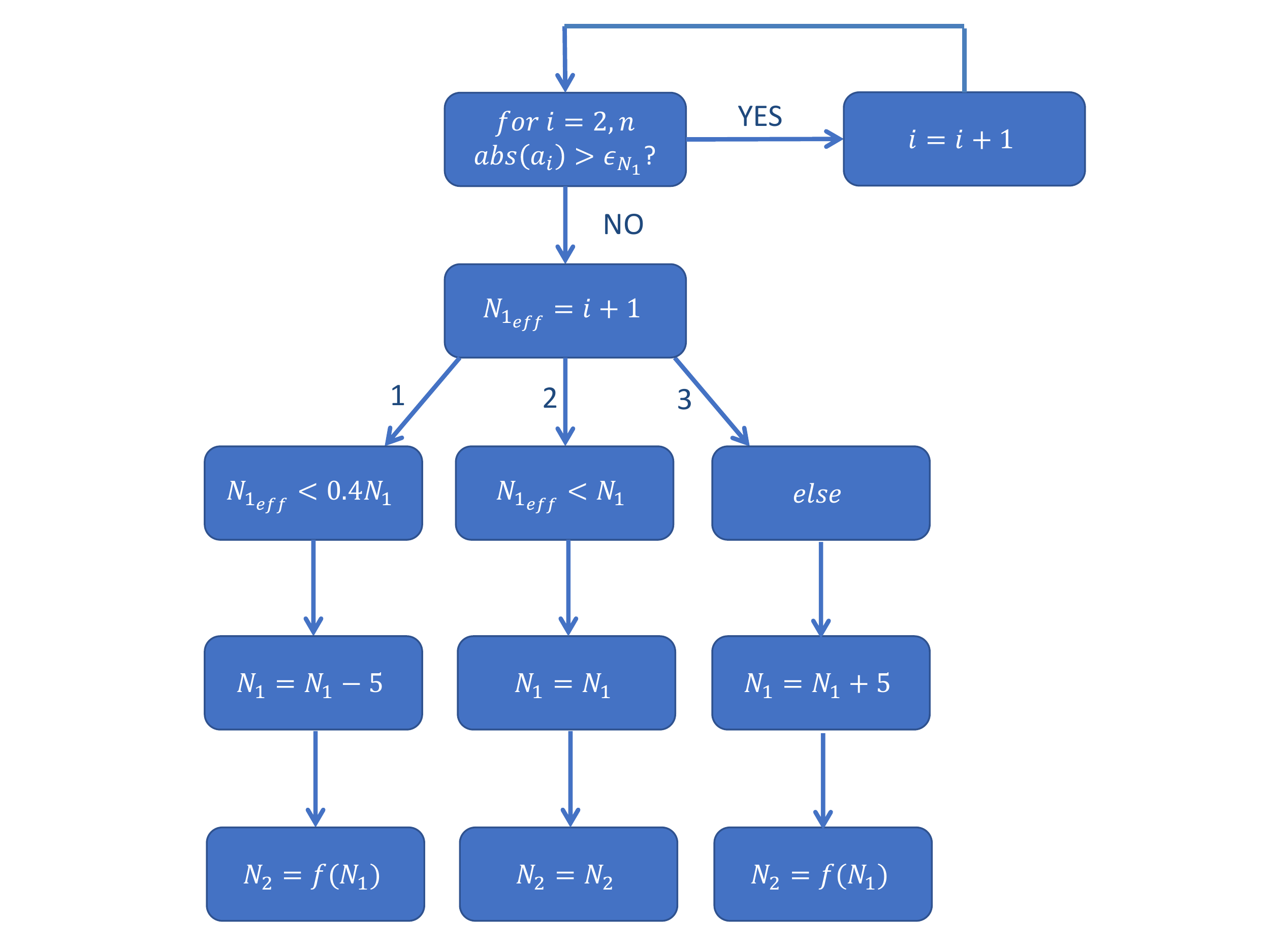}
	\caption{Automatic calculation of $N_1$ and $N_2$ performed in CN-Stream.}
	\label{fig:auto_N1}
\end{figure}

The procedure is depicted in Fig. \ref{fig:auto_N1} and follows the main steps:
\begin{itemize}
	\item Solve the problem with an initial set of values for $N_1$ and $N_2$
	\item Look at the modal amplitudes $a_n$ deducing the efficient number of modes $N_{1_{eff}}$ satisfying $abs(a_{N_{1_{eff}}})<\epsilon_{N_1}$. Then, three configurations possible:
\begin{itemize}
\item if $N_{1_{eff}} < 0.4N_1$, then $N_1$ is decreased by 5,
\item else if $N_{1_{eff}} < N_1$ then nothing is done,
\item else $N_1$ is increased by 5.
\end{itemize}
	\item If $N_1$ changed, the number of modes $N_2$ of the elevation is deduced from $N_1$ by an empirical formula calibrated in section \ref{N1_N2}:
\begin{equation}
N_2 \egal N_1 \left(1.5 + \frac{1.5}{0.3}\max \left|\frac{\partial \eta}{\partial x}\right|\right).
\end{equation}
\end{itemize}

The procedure is stopped when the solution of the problem is achieved at the target accuracy (iterative solution of the linear system) and when the number of modes is unchanged in the previous algorithm, meaning it is optimal for the current configuration.

%The algorithm is presented in Fig.\ref{fig:auto_N1} and consists firstly in finding the first value of $n$, from $n=N_1$ to $n=2$, that satisfies as:
%\begin{equation}
%abs(a_n) > \epsilon_{N_1},
%\end{equation} 
%where $\epsilon_{N_1}$ represents the convergence criteria on the modes (option\%eps\_N1).
%Noting $i$ the number of iterations needed to find the threshold, we affect:
%\begin{equation}
%N_{1_{eff}} = i+1.
%\end{equation} 

\section{Results}
\label{Results}

This section presents different results obtained with the CN-Stream code. The objective of this part is to detail the numerical properties of the method and especially to demonstrate the relevance of the enhancements proposed. The highest waves accessible with the current method are also provided explicitly as a matter of completeness.

In addition, different applications of the CN-Stream model to the study of non-linear regular waves are presented. In the text some references are done to the parameters names in the input files, which are further described in section \ref{subsubsec:Input}.

%\subsection{Some examples: Modes $a_n$ and $b_n$}
\subsection{Some examples: Modal description of quantities}
In this paragraph three different wave conditions are simulated corresponding respectively to infinite, finite and shallow water depths. The corresponding wave parameters are given in Tab.\ref{tab:exemple}.
For each wave condition, the elevation and the slope are presented as a function of the phase $kx$, as well as the modal amplitudes of the elevation and velocity potential. Then, the maximal steepnesses available for different water depths are presented.

\begin{table}[!htbp]
	\centering
\begin{tabular}{llllllll}
\toprule
& Period (s) & Height (m) & Depth (m) & $H/gT^2$ & $kH$ & $h/gT^2$ & $kh$ \\
\midrule 
Infinite water depth & 8  & 15 & Inf & 0.024  & 0.80 & Inf   & Inf\\
Finite water depth   & 8  & 14 & 37  & 0.022  & 0.78 & 0.059 & 2.0 \\
Shallow water depth  & 25 & 10 & 37  & 0.0016 & 0.13 & 0.006 & 0.48\\
\bottomrule
\end{tabular}	\caption{Wave parameters for the three studied conditions.}
	\label{tab:exemple}
\end{table}
\subsubsection{Infinite water depth}\label{sec:ex-prof-inf}
\begin{figure}[!htbp]
	\centering
		\includegraphics[width=0.48\textwidth]{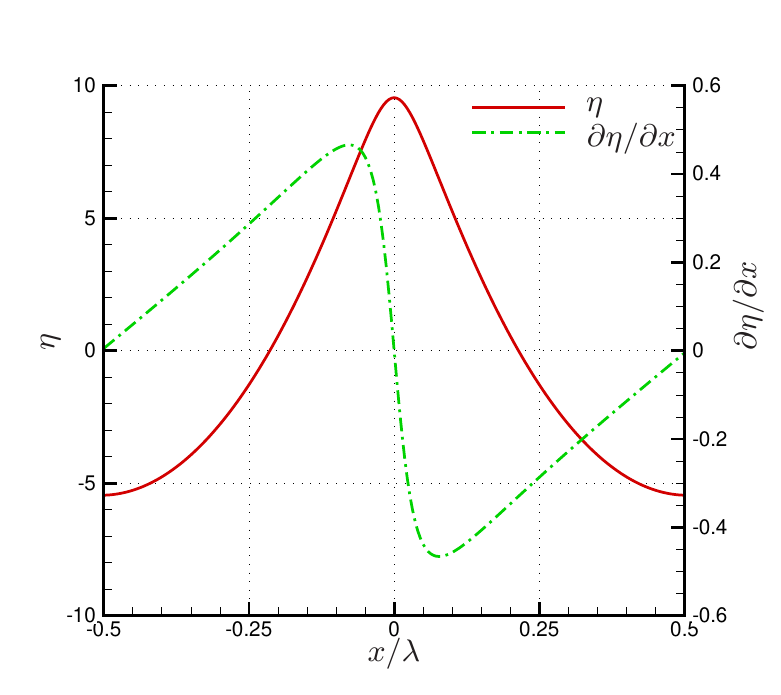}
		\includegraphics[width=0.48\textwidth]{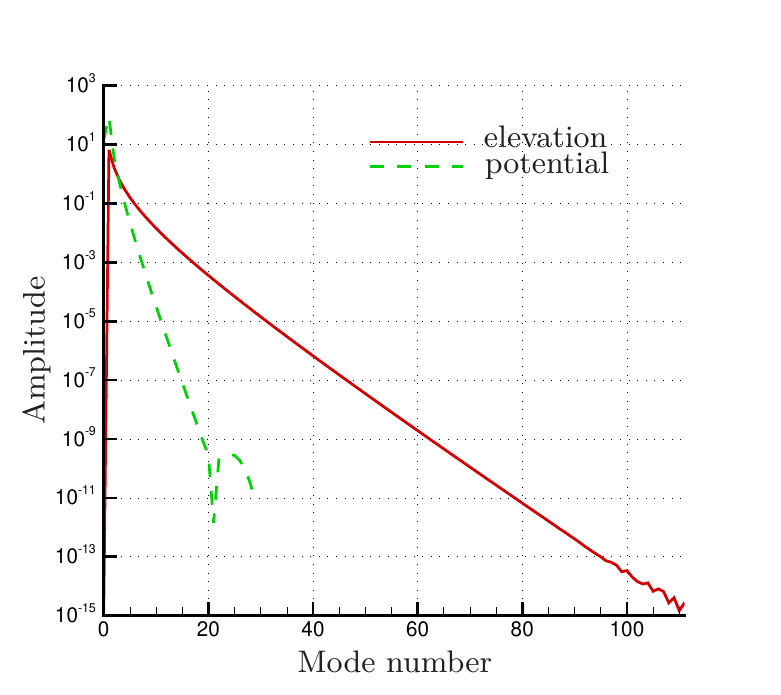}
%		\psfragfig*[width=0.48\textwidth]{Figures/RF_space_T8_H15_hInf_new}{\small
%		\psfrag{a}[l][]{$\eta$}
%		\psfrag{b}[l][b]{$\partial \eta/\partial x$}
%		\psfrag{x}[][]{$x/\lambda$}
%		}
%		\psfragfig*[width=0.48\textwidth]{Figures/RF_modes_T8_H15_hInf_new}{\small
%		\psfrag{a}[l][t]{elevation}
%		\psfrag{b}[l][]{potential}
%		\psfrag{k}[t][]{Mode number}
%		\psfrag{c}[b][]{Amplitude}
%		}
	\caption{Infinite water depth -  wave period $T=8$s and wave height $H=15$m. Left: Wave surface elevation and slope. Right: Modal amplitudes of the surface elevation and of the velocity potential.}
	\label{fig:T8H15hInf}
\end{figure}

In Fig.\ref{fig:T8H15hInf}, an example of a wave propagating over an infinite water depth with a wave period $T=8$s and a wave height $H=15$m is presented.  The wave surface elevation and the slope are shown as well as the modal amplitudes of the free surface elevation $\eta$ and the velocity potential $\phi$.\\

%Note: $\epsilon_{N_1}=10^{-12}$ (relative)

For such high steepness $kH=0.80$, the well-known non-linear features of the free surface elevation are recovered, namely a strong asymmetry between the crest and the trough, together with large value of the local steepness.

It is also clear from the modal description that the necessary number of modes is different for $\eta$ and $\phi$ due to a different convergence rate of the modal amplitudes. Thanks to the proposed enhanced algorithm, one can reach an accuracy on the amplitude of the mode of $\epsilon_{N_1}=10^{-12}$ (defined as relative error).\\

\begin{figure}[!htbp]
	\centering
	\includegraphics[width=0.48\textwidth]{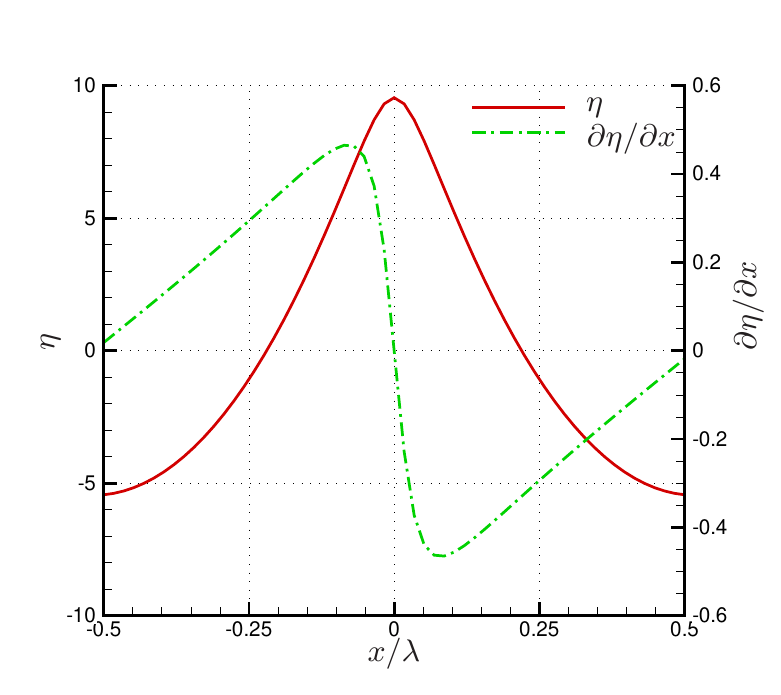}
	\includegraphics[width=0.48\textwidth]{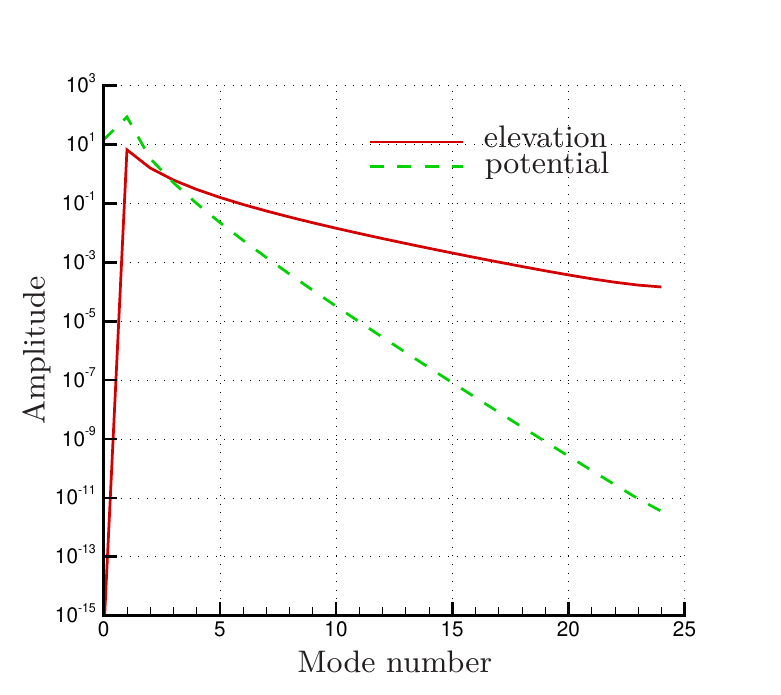}
%	\psfragfig*[width=0.48\textwidth]{Figures/RF_space_T8_H15_hInf_N1eqN2}{\small
%		\psfrag{a}[l][]{$\eta$}
%		\psfrag{b}[l][b]{$\partial \eta/\partial x$}
%		\psfrag{x}[][]{$x/\lambda$}
%		}
%		\psfragfig*[width=0.48\textwidth]{Figures/RF_modes_T8_H15_hInf_N1eqN2}{\small
%		\psfrag{a}[l][t]{elevation}
%		\psfrag{b}[l][]{potential}
%		\psfrag{k}[t][]{Mode number}
%		\psfrag{c}[b][]{Amplitude}
%		}
	\caption{Infinite water depth -  wave period $T=8$s and wave height $H=15$m. Non-optimal CN-Stream solution with $N_1=N_2$. Left: Wave surface elevation and slope. Right: Modal amplitudes of the surface elevation and of the velocity potential.}
	\label{fig:T8H15hInf_N1eqN2}
\end{figure}

As a matter of comparison to the original stream function model \cite{Rienecker}, the same algorithm is applied, fixing the same number of modes for the two quantities (\emph{i.e.} $N_1=N_2$). The results are depicted in Fig. \ref{fig:T8H15hInf_N1eqN2}.

The free surface looks the same than previously in the spatial domain, but even if the modal description of $\eta$ and $\phi$ are still convergent, the level of accuracy is reduced compared to the enhanced stream function model. The results of Fig. \ref{fig:T8H15hInf_N1eqN2} are actually the highest accuracy (\emph{i.e.} smallest amplitude of highest mode) one can possibly reach when using $N_1=N_2$. The amplitude of the smallest mode for the decription of the free surface elevation is now $\epsilon=2~ 10^{-5}$ to compare with $\epsilon_{N_1}=10^{-12}$ in the previous configuration.

\begin{figure}[!htbp]
	\centering
	\includegraphics[width=0.48\textwidth]{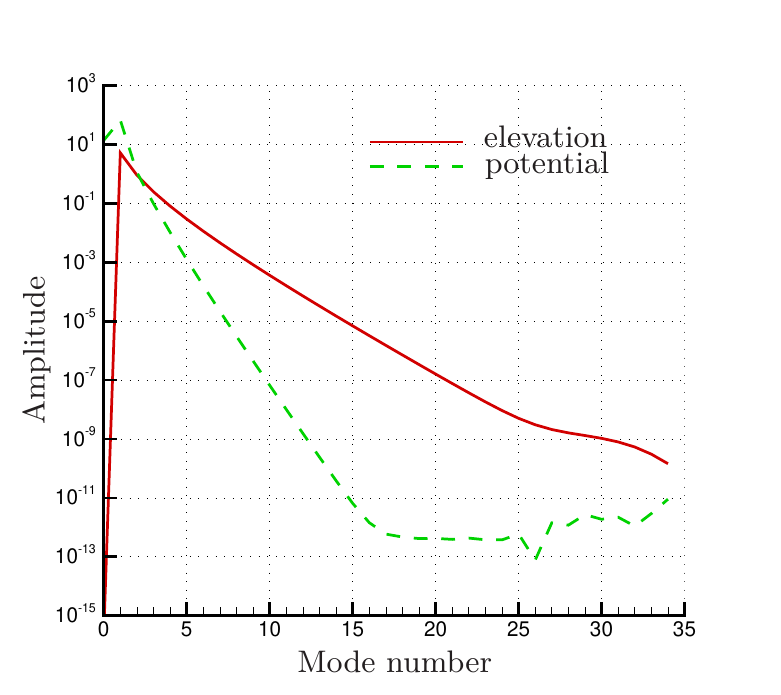}
%		\psfragfig*[width=0.48\textwidth]{Figures/RF_modes_T8_H11_hInf_N1eqN2}{\small
%		\psfrag{a}[l][t]{elevation}
%		\psfrag{b}[l][]{potential}
%		\psfrag{k}[t][]{Mode number}
%		\psfrag{c}[b][]{Amplitude}
%		}
	\caption{Infinite water depth -  wave period $T=8$s and wave height $H=11$m. Non-optimal CN-Stream solution with $N_1=N_2$. Modal amplitudes of the surface elevation and of the velocity potential.}
	\label{fig:T8H11hInf_N1eqN2}
\end{figure}

The accuracy is actually limited by the fact that if one increases the number of modes for the description of $\eta$, the consequent increase in the description of $\phi$ may create some numerical instabilities. As an example, Fig. \ref{fig:T8H11hInf_N1eqN2} depicts the initiation of such process for a regular wave in infinite depth with a smaller wave height ($T=8$s and wave height $H=11$m).

For this wave steepness, one can reach a relative amplitude of the smallest mode $\epsilon=5~10^{-11}$. It is clearly seen that the decrease of the modal amplitudes of the velocity potential reach a plateau after the mode number $16-17$. These highest modes, which do not decrease in amplitude any more are responsible of the enhanced behaviour observed of CN-Stream compared to original implementation of \cite{Rienecker}. This comes from the involved spatial derivatives of the quantities, corresponding to a multiplication by $k$ in the modal space that will induce a non convergent Fourier description of the corresponding quantity.

\subsubsection{Finite water depth ($h'=h/gT^2=0.059$ and $kh=2.0$)}
\begin{figure}[!htbp]
	\centering
	\includegraphics[width=0.48\textwidth]{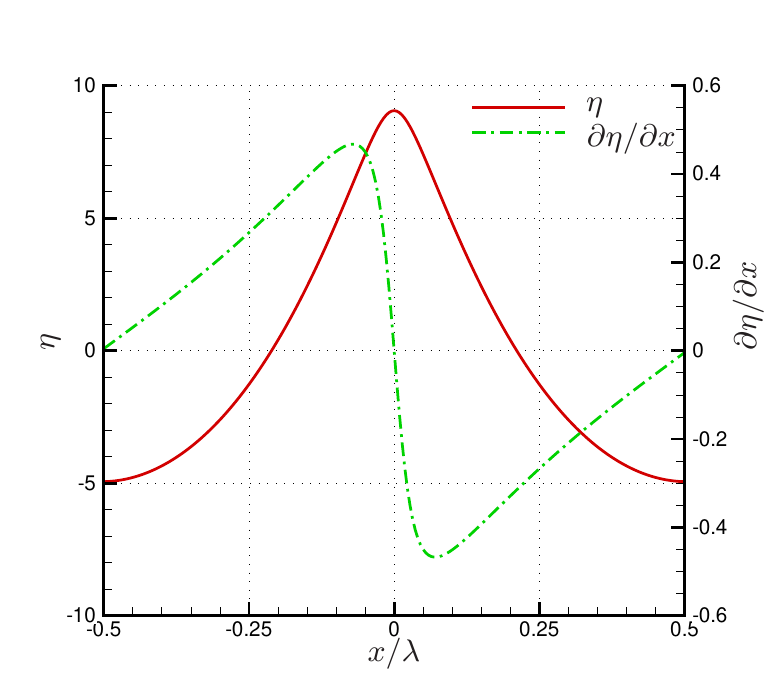}
	\includegraphics[width=0.48\textwidth]{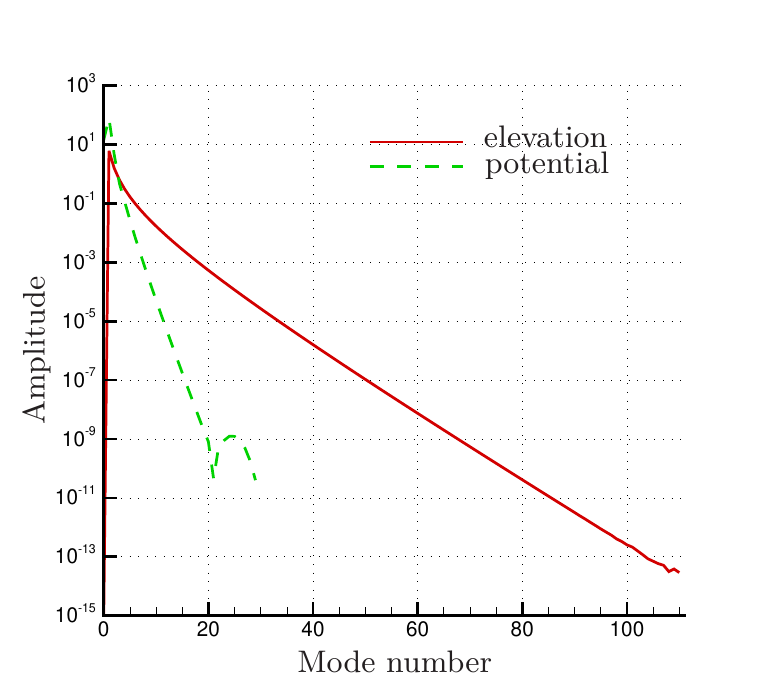}
%		\psfragfig*[width=0.48\textwidth]{Figures/RF_space_T8_H14_h37_new}{\small
%		\psfrag{a}[l][]{$\eta$}
%		\psfrag{b}[l][b]{$\partial \eta/\partial x$}
%		\psfrag{x}[][]{$x/\lambda$}
%		}
%		\psfragfig*[width=0.48\textwidth]{Figures/RF_modes_T8_H14_h37_new}{\small
%		\psfrag{a}[l][t]{elevation}
%		\psfrag{b}[l][]{potential}
%		\psfrag{k}[t][]{Mode number}
%		\psfrag{c}[b][]{Amplitude}
%		}
%		%\includegraphics[width=0.49\textwidth]{Figures/RF_space_T8_H14_h37.eps}
%		%\includegraphics[width=0.49\textwidth]{Figures/RF_modes_T8_H14_h37.eps}
	\caption{Finite water depth -  wave period $T=8$s, wave height $H=14$m and water depth $h=37$m. Left: Wave surface elevation and slope. Right: Modal amplitudes of the surface elevation and of the velocity potential.}
	\label{fig:T8H14h37}
\end{figure}

In Fig.\ref{fig:T8H14h37}, an example of a non-linear regular wave propagating over a finite water depth ($h=37$m) with a wave period $T=8$s and a wave height $H=14$m is presented.  The wave surface elevation and the slope are shown as well as the modal amplitudes of $\eta$ and $\phi$.

This confiugration is usually known as intermediate water depth ($kh=2.0$) with consequently limited effect of the presence of the sea floor. Comparing with Fig. \ref{fig:T8H15hInf}, the results are similar with a a free surface profile exhibiting slightly longer troughs and a modal description requiring similar number of modes ($N_1=30$ and $N_2=111$) to reach the same accuracy $\epsilon_{N_1}=10^{-12}$.
\subsubsection{Shallow water depth ($h'=h/gT^2=0.006$ and $kh=0.48$)}
\begin{figure}[!htbp]
	\centering
	\includegraphics[width=0.48\textwidth]{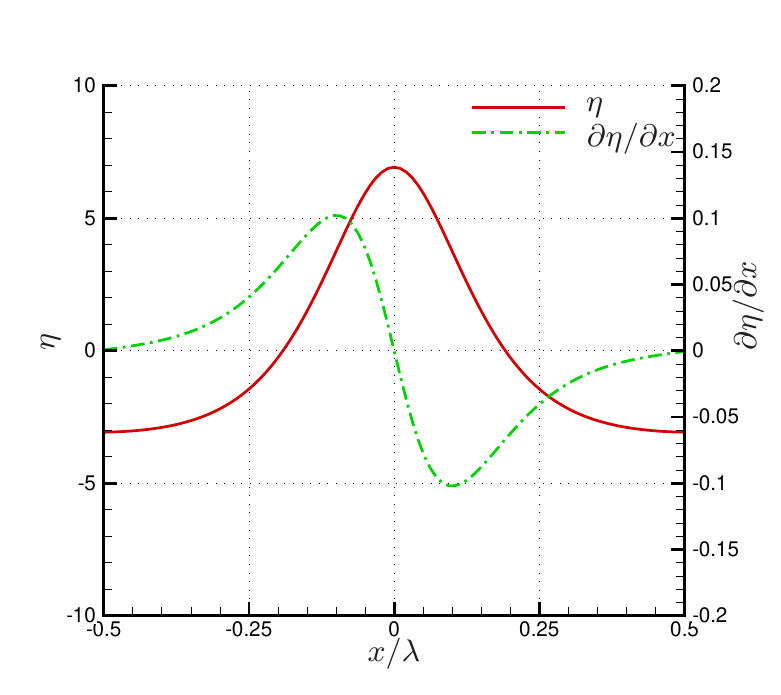}
	\includegraphics[width=0.48\textwidth]{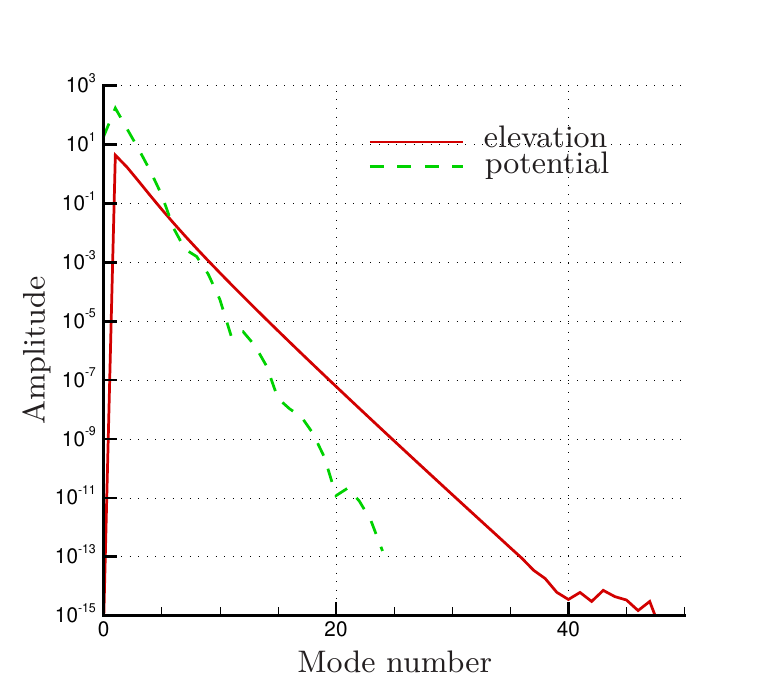}
%		\psfragfig*[width=0.48\textwidth]{Figures/RF_space_T25_H10_h37_new}{\small
%		\psfrag{a}[l][]{$\eta$}
%		\psfrag{b}[l][b]{$\partial \eta/\partial x$}
%		\psfrag{x}[][]{$x/\lambda$}
%		}
%		\psfragfig*[width=0.48\textwidth]{Figures/RF_modes_T25_H10_h37_new}{\small
%		\psfrag{a}[l][t	]{elevation}
%		\psfrag{b}[l][]{potential}
%		\psfrag{k}[t][]{Mode number}
%		\psfrag{c}[b][]{Amplitude}
%		}
%		%\includegraphics[width=0.49\textwidth]{Figures/RF_space_T25_H10_h37.eps}
%		%\includegraphics[width=0.49\textwidth]{Figures/RF_modes_T25_H10_h37.eps}
	\caption{Finite water depth -  wave period $T=25$s, wave height $H=10$m and water depth $h=37$m. Left: Wave surface elevation and slope. Right: Modal amplitudes of the surface elevation and of the velocity potential.}
	\label{fig:T25H10h37}
\end{figure}

The last example in this part deals with a wave propagating over a the same water depth than previous one ($h=37$m) but with a significantly longer wave period $T=25$s. This corresponds to a shallow water wave configuration ($kh=0.48$) and a wave height $H=10$m. The wave surface elevation and the slope are shown as well as the modal amplitudes of $\eta$ and $\phi$ in Fig.\ref{fig:T25H10h37}.

The physical effects associated to the shallowness of the water depth are now clear in this configuration with very clear assymetries between crest and trough in the horizontal and the vertical directions. In terms of modal representation, it is interesting to note that when going to shallower water depth, the decrease rate of the modal amplitudes of $\eta$ and $\phi$ becomes closer one with the other. As a consequence, the necessary number of points to reach the target accuracy $\epsilon_{N_1}=10^{-12}$ is now $N_1=25$ and $N_2=49$.
%
%
%
%\subsection{On the choice of the modes}
\subsection{Limiting waves}
\label{subsubsec:limiting_waves}

It appears interesting for the user to have an idea of the waves that can be computed with CN-Stream. We remind that the important physical parameters are the steepness  $kH$ and the relative water depth $kh$ (or a combination of these two parameters such as height to depth ratio $H/h$ or the Ursell number $Ur=\frac{H\lambda^2}{h^3}=4\pi^2\frac{kH}{(kh)^3}$).
\newline

The results are dependent of the numerical parameters. The following parameters are used in the present section (in brackets the corresponding input file option, see section \ref{programDoc} for details):
\begin{itemize}
	\item $n_H = 100$ (option: n\_H)
	\item Relative error (option: err\_type = 1)
	\item $\epsilon^{rel}_F = 10^{-10}$ (option: eps\_err) 
	\item $\max(\epsilon^{rel}_F)   = 10.0$ (option: err\_max) 
	\item $\epsilon^{rel}_Z = 10^{-10}$  (option: eps\_inc)
	\item $\epsilon_{N_1}   = 10^{-10}$
\end{itemize}

%\begin{itemize}
%	\item option\%n\_H=100
%	\item option\%err\_type = 1: Relative error
%	\item option\%eps\_err  = 1d-10 
%	\item option\%err\_max  = 10.0d0 
%	\item option\%eps\_inc  = 1d-10
%	\item option\%eps\_N1   = 1d-10
%\end{itemize}

Moreover, computations are performed in non-dimensional form (waveInput: GeneralDimension = 0 see section \ref{programDoc}) and without any eulerian current (input: CurrentType=0 and input: CurrentValue=0.0). Tables \ref{tab:applicabilite_lambda} \& \ref{tab:applicabilite_periode}  present the numerical values obtained for the different tests that have been performed. $H_{lim}$ stands for the heighest wave accessible with CN-Stream in the configuration tested.
 
\begin{table}[htbp]
   \centering
   \begin{tabular}{@{}lccccccc@{}}
      \toprule
       \multicolumn{8}{c}{Input  is $\lambda$ (input: WaveInput=1 )}  \\
      \midrule
      Relative water depth $kh$ & $\infty$ & 3.0  & 1.0 & 0.5 & 0.3 & 0.2 & 0.1 \\
      \addlinespace
      Limiting steepness $kH_{lim}$ & 0.84 & 0.84 & 0.60 & 0.34 & 0.22 & 0.15 & 0.077 \\
       \addlinespace
       $H_{lim}/h$ & undef. & 0.28 & 0.60 & 0.68 & 0.72 & 0.74 & 0.77 \\
      \bottomrule
   \end{tabular}
   \caption{Maximal steepnesses that can be computed for different water depths - non-dimensional input is the wavelength $\lambda$.}
	\label{tab:applicabilite_lambda}
\end{table}
  
\begin{table}[htbp]
\centering
    \begin{tabular}{@{}lccccccc@{}}
      \toprule
       \multicolumn{8}{c}{Input is $T$ (input: WaveInput=0)}  \\
      \midrule
      Relative water depth $h/(gT^2)$ & $\infty$ & 0.1  & 0.02 & 0.005 & 0.002 & 0.001 & 0.0005 \\
      \addlinespace
      Limiting steepness $H_{lim}/(gT^2)$ & 0.025 & 0.025 & 0.012 & 0.0035 & 0.0015 & 0.00075 & 0.00038 \\
       \addlinespace
       $H_{lim}/h$ & undef. & 0.25 & 0.61 & 0.70 & 0.73 & 0.75 & 0.76 \\
      \bottomrule
   \end{tabular}
   \caption{Maximal steepnesses that can be computed for different water depths - non-dimensional input is the period $T$.}
	\label{tab:applicabilite_periode}
\end{table}

\begin{figure}[h!tbp]
	\centering
	\includegraphics[width=\textwidth]{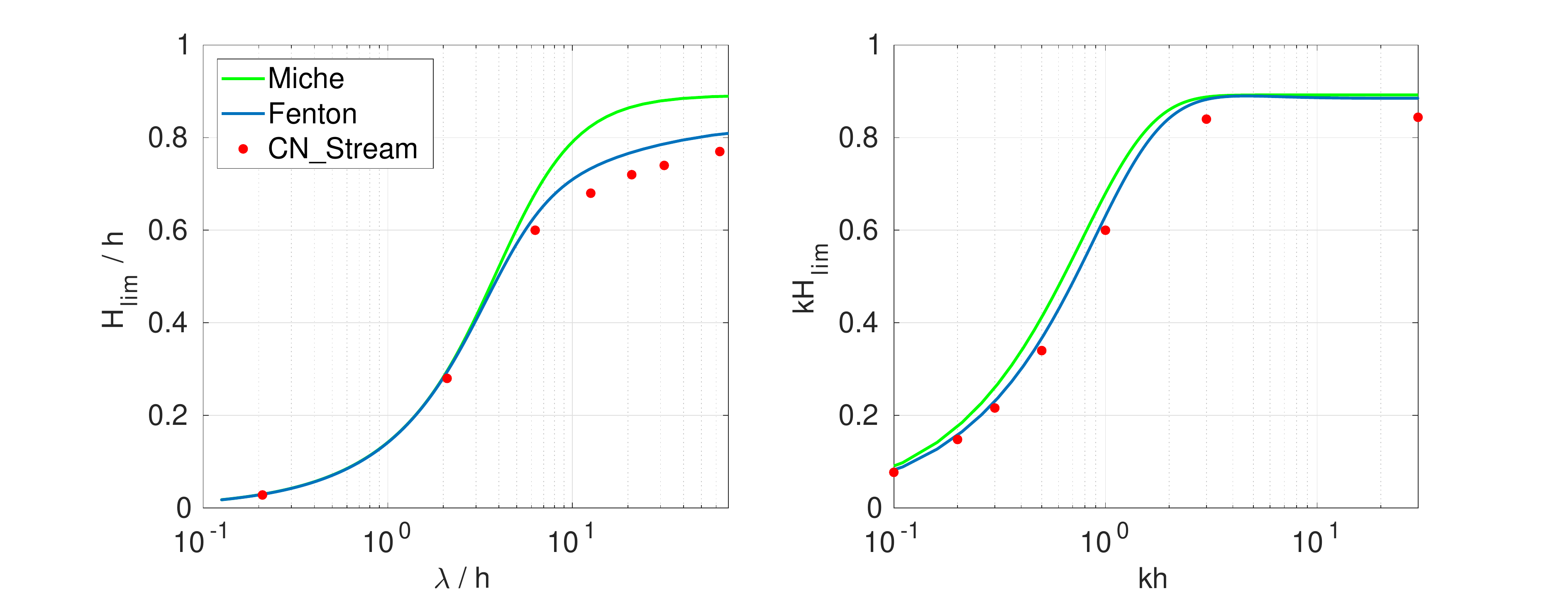}
	\caption{The region in which solutions for steady waves can be obtained with CN-Stream (dots representing the highest wave accessible $H_{lim}$ for given input parameters). Comparison to the theoretical formulas of \cite{Miche1944b} and \cite{fenton1990nonlinear}. Input is the wavelength. }
	\label{fig:applicabilite_lambda}
\end{figure}

\begin{figure}[h!tbp]
	\centering
	\includegraphics[width=0.5\textwidth]{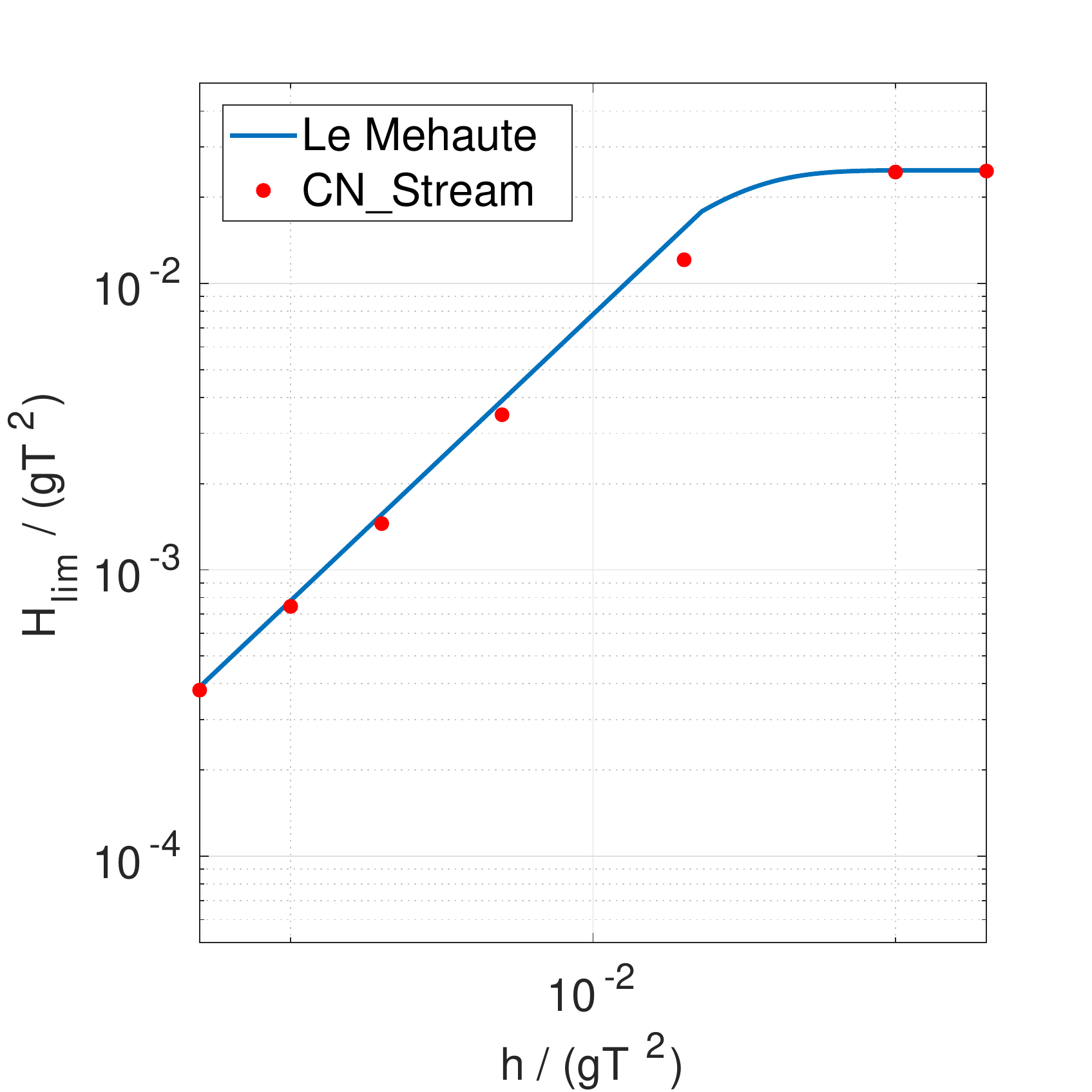}
	\caption{The region in which solutions for steady waves can be obtained with CN-Stream (dots representing the highest wave accessible $H_{lim}$ for given input parameters). Comparison to the theoretical formulas of \cite{le_mehaute_1976}. Input is the period.}
	\label{fig:applicabilite_T}
\end{figure}

In order to be clearer and to compare the results presented in Tabs. \ref{tab:applicabilite_lambda} \& \ref{tab:applicabilite_periode} to the theoretical formulas of limiting regular waves in various conditions, the preceding results are plotted in Figs. \ref{fig:applicabilite_lambda} \&  \ref{fig:applicabilite_T}.\\

Limits to the existence of waves have been first parametrized by \cite{Miche1944b}. He proposed a simple formula for the maximal steepness that can be computed given by:
\begin{equation}
\epsilon_{lim}=H_{lim}/\lambda,
\label{eq:Miche}
\end{equation}
for a large range of depths $h$. This equation \eqref{eq:Miche} has been validated thanks to experimental and numerical data and takes now the following form: 
\begin{eqnarray}
\label{eq:limit_regular_Miche}
\epsilon_{lim}=0.142 \tanh \left( kh \right).
\end{eqnarray}

It should be noted that in very shallow water depths ($kh \rightarrow 0$), this equation \eqref{eq:limit_regular_Miche} overestimates the maximum computed height ($H_{lim}/h \rightarrow 2 \pi * 0.142 = 0.892$). Various studies have tried to improve this simple formula. For instance \cite{Williams1981} studied experimentally the gravity waves stability in a large range of relative water depths. \cite{fenton1990nonlinear} used the experimental results to propose a parametrized formula under the form:
\begin{eqnarray}\label{eq:limit_regular}
\frac{H_{lim}}{h} = \frac{0.141063 \left(\frac{\lambda}{h}\right) + 0.0095721 \left(\frac{\lambda}{h}\right)^2 + 0.0077829 \left(\frac{\lambda}{h}\right)^3}{1+0.0788340 \left(\frac{\lambda}{h}\right) + 0.00317567 \left(\frac{\lambda}{h}\right)^2 + 0.0093407 \left(\frac{\lambda}{h}\right)^3}.
\end{eqnarray}
This formula presents the advantages to accurately treat the following limiting cases:
\begin{itemize}
\item infinite depth and $kH_{lim}=0.885$,
\item solitary wave in very shallow water depth $H_{lim}/h=0.833$ (see for instance \cite{hunter1983solitary})
\end{itemize}  
The two preceding equations \eqref{eq:limit_regular_Miche} \& \eqref{eq:limit_regular} use non-dimensional quantities with respect to the wavelength $\lambda$. It can also be useful to non-dimensionalize the quantities by the period, as presented in Le M\'ehaut\'e's diagram (see Fig.~\ref{fig:Le_Mehaute}, Fig.~\ref{fig:applicabilite_T} and \cite{le_mehaute_1976}). This diagram presents the limit in terms of wave height $H_{lim}/(gT^2)$ as function of the relative water depth $h/(gT^2)$.\\

As a summary, with the chosen high level of accuracy, those results demonstrate that the CN-Stream code allows the simulation of non-linear regular waves up to waves close to the breaking limit. If one intends to simulate even higher waves, the acceptable level of error needs to be reduced.

\subsection{Nonlinear effects}
\label{subsec:Nonlinear_effects}

This section is dedicated to the study of some of the non-linear features associated to regular water waves. These are useful in the definition of some properties for the numerical solution.

\subsubsection{Influence on the wavelength}
\begin{figure}[h!tbp]
	\centering
		\includegraphics[width=0.49\textwidth]{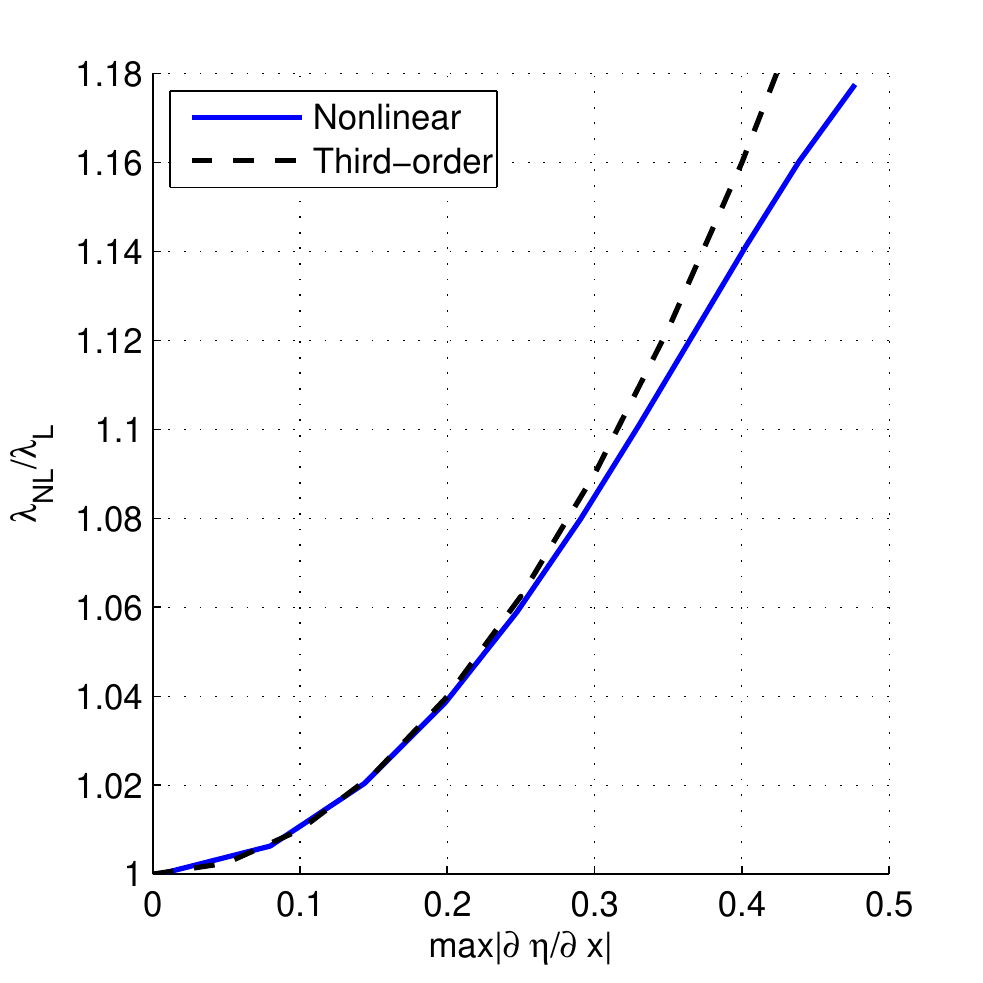}
		\includegraphics[width=0.49\textwidth]{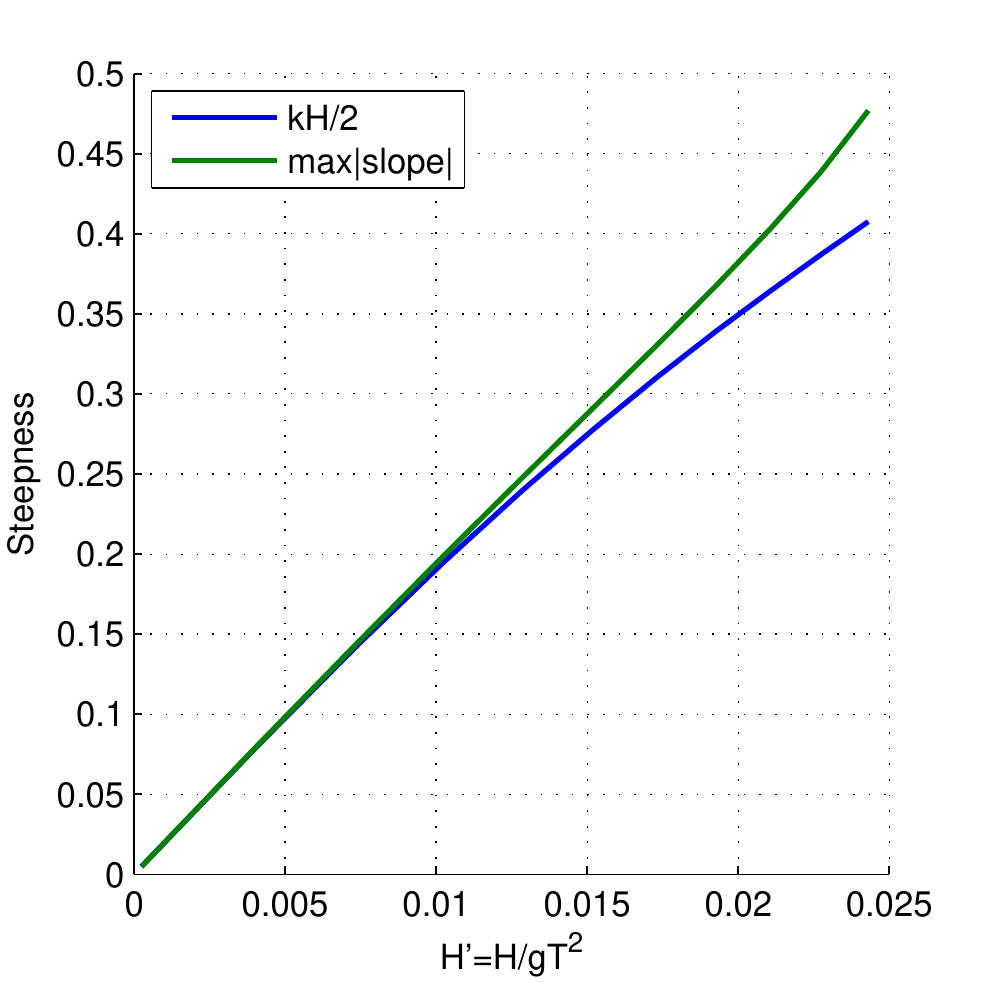}
	\caption{Wavelength (left) and steepness (right) evolutions for a wave propagating over an infinite water depth.}
	\label{fig:eps-k-hInf}
\end{figure}

\paragraph{Infinite water depth}
In infinite water depth, the only non-dimensional parameter characterizing the wave is the steepness. Figure \ref{fig:eps-k-hInf} (left) shows the evolution of the wavelength with the "real" slope (measured as the maximum of the slope  $|\partial \eta/\partial x|$ over the wavelength). A good agreement is found with the third-order formula:
\begin{equation}
\frac{\lambda_{NL}}{\lambda_{L}} \egal 1 + (ka)^2,
\end{equation}
with $ka=\frac{kH}{2}=\max |\partial \eta/\partial x|$, until $ka \simeq 0.3$.

The evolution of the two different definitions for the steepness ($kH/2$ and maximum slope) as a function of the non-dimensional height $H'=H/gT^2$ is also provided in Fig. \ref{fig:eps-k-hInf}. It appears that the steepness defined as the maximum slope as an almost linear dependence with $H'$ over the whole range of existence of the wave (except for the most extreme ones), while $kH/2$ exhibits a more complex evolution, which is linear only for waves with moderate steepness.

\begin{figure}[h!tbp]
	\centering
		\includegraphics[width=0.5\textwidth]{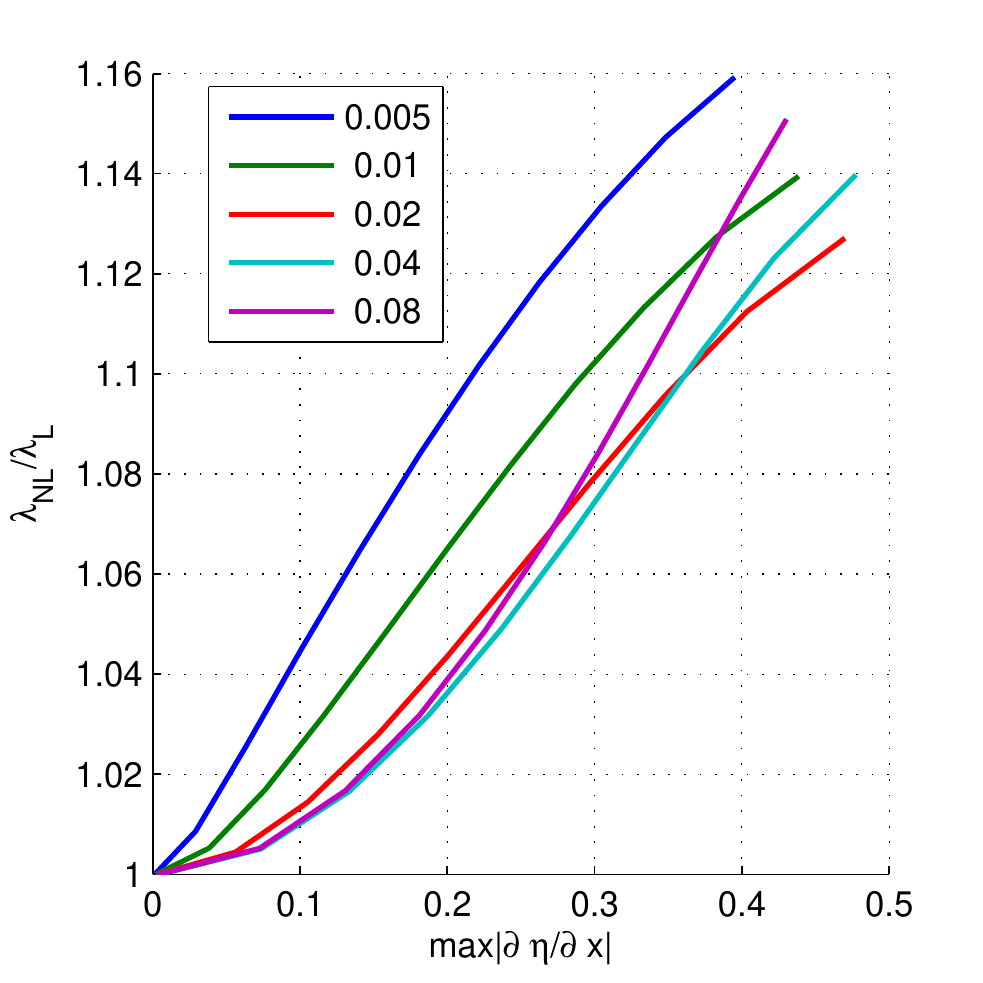}
	\caption{Non-linear evolution of the wavelength for a wave propagating in finite water depths (the legend gives the values of $h'=h/gT^2$).}
	\label{fig:lambda-hvar}
\end{figure}

\paragraph{All water depths}
Then, Fig.\ref{fig:lambda-hvar} shows the non-linear evolution of the wavelength as a function of the maximal wave slope. 
The  whole range of depths is covered from shallow water depths to infinite water depths, as shown in Tab. \ref{tab:kh}.
 
\begin{table}[h!tbp]
	\centering
\begin{tabular}{llllll}
\toprule
$k_Lh$ & 0.2   & 0.4  & 0.8  & 1.6  & 3.2\\
$h'=h/gT^2$ & 0.005 & 0.01 & 0.02 & 0.04 & 0.08\\
\bottomrule
\end{tabular}	\caption{Non-dimensional depths.}
	\label{tab:kh}
\end{table}

For small slopes, the increase of the wavelength is more important for small relative water depths. For larger slopes, the modification of the wavelength does not exhibit a specific trend with the relative water depth anymore, even if the shallower water depth seems to always exhibit the largest increase in non-linear wave length.

Note that depending on the relative water depth, the maximum slope observed for the steepest wave computed is varying in the range $\max \left| \partial \eta/\partial x\right| \in \left[ 0.40 ; 0.48 \right]$. Similarly, we observe that the maximal modification in wave length is in a small range $\left[ 13 \% ; 16 \%\right]$.

The non-linear modification of the wavelength is consequently moderate. There is thus no explicit need to use the non-linear wavelength when computing the non-dimensional parameters such as $kh$ and $kH$.
\newline

\begin{figure}[!htbp]
	\centering
		\includegraphics[width=0.7\textwidth]{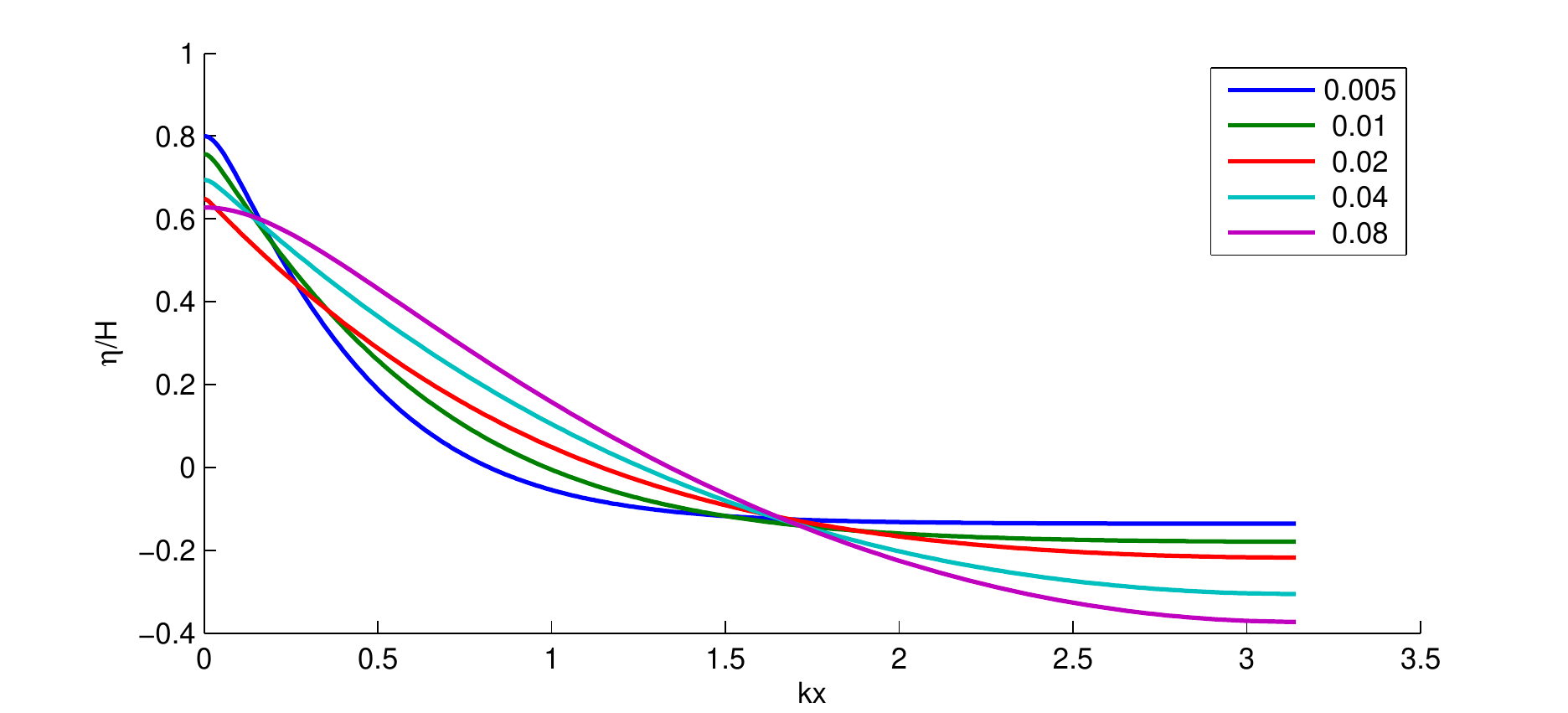}
	\caption{Wave elevation obtained for a maximal slope (the legend gives the values of $h'=h/gT^2$).}
	\label{fig:eta-hvar}
\end{figure}

Figure \ref{fig:eta-hvar} presents the wave elevation obtained for the maximal slope at various water depths.

As expected, the crest-trough asymetry is enhanced when reducing the relative water depth (both in terms of amplitude and relative length). The numerical solution of CN-Stream in small water depth recovers the cnoidal wave features.

%On peut donner l'amplitude des 3 premiers modes, en normalisant pour faire appara�tre l'�cart � la solution de Stokes en profondeur infinie. Attenion toutefois, � partir du troisi�me ordre, les �changes d'�nergie sont permis et les instabilit�s apparaissent
%
%
%
\subsubsection{Maximal slope}
\label{slope}
\paragraph{Infinite water depth}
It is interesting to compare the various definitions of the steepness (the linear steepness $kH/2$ and the maximal slope)  as a function of the relative wave height $H'=H/gT^2$.  The following relationship is expected:
\begin{equation}
H' \egal \frac{H}{gT^2} \egal \frac{k_L H}{2} \, \frac{1}{2\pi^2}
\end{equation}
with $2\pi^2\simeq 19.7$. 
 From Fig.\ref{fig:eps-k-hInf} (right) we observe that $\max |\partial \eta/\partial x|=19 H'$ for all $H'$. It means that $H'$ is a very good measurement of the wave slope non-linearity for the infinite water depth case. The linear steepness $kH/2$ is moving away from the maximal wave slope as soon as  $H'>0.015$.

\begin{figure}[!htbp]
	\centering
	\includegraphics[width=0.45\textwidth]{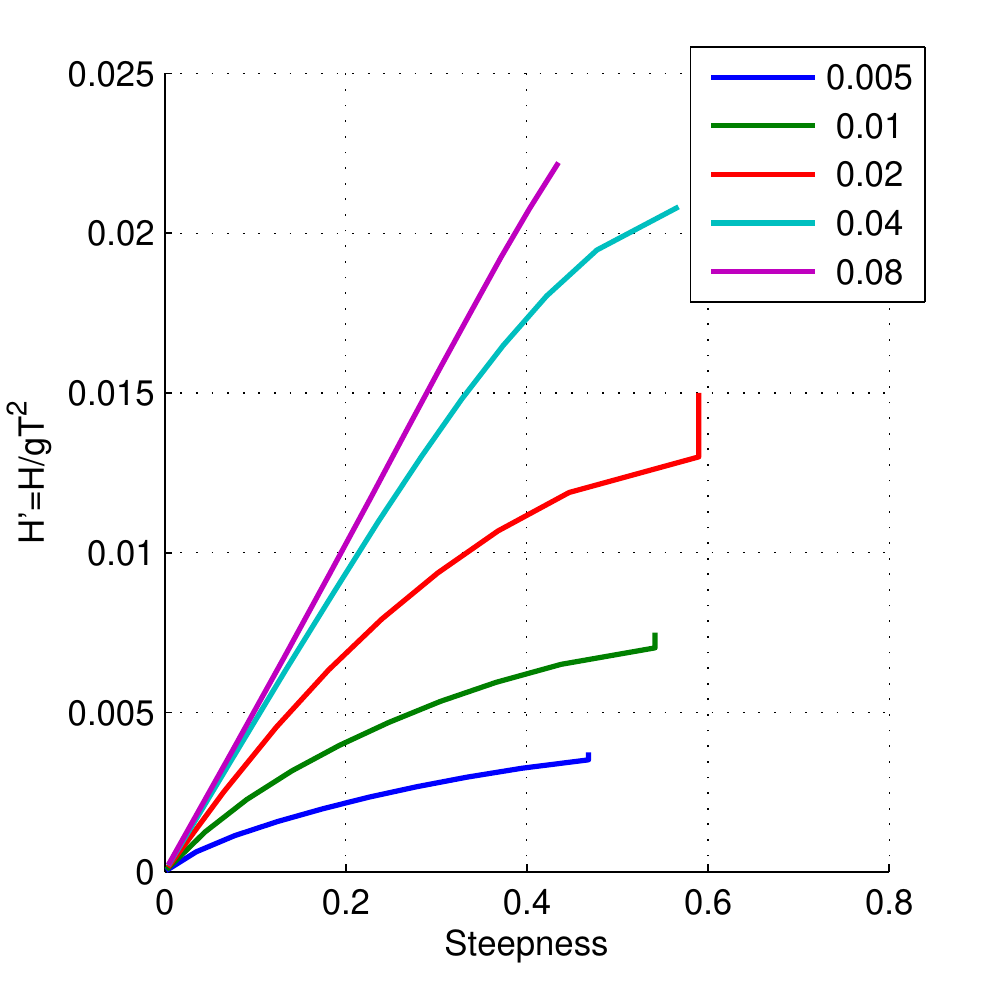}
	\caption{Steepnesses for a wave propagating over a finite water depth (the legend gives the values of $h'$). }
	\label{fig:eps-hvar}
\end{figure}

\paragraph{All water depths} 
The evolution between the slope and $H'$ is presented in Fig.\ref{fig:eps-hvar} for different water depths.
One can observed that the relative water depth $k_Lh=3$ ($h'=0.08$) already corresponds to the infinite water depth: results for larger water depths are superimposed to those obtained at $k_Lh=3$. This corresponds to the usual definition of waves considered as deep-water when $h/\lambda > 0.5$. 

For an infinite water depth (see paragraph above), we observed that $H'$ was proportional to the maximal wave slope for all wave steepnesses. 
Here, Fig.~\ref{fig:eps-hvar} shows that for a shallow water depth, the relationship between  $H'$  and the steepness is linear only for small slopes. Thus the parameter  $H'$ is not a good measurement of the wave non-linearity in shallow water. Indeed, if the value of $H'$ is multiplied by 2, the maximal wave slope is multiplied by a factor larger than 2, showing that the wave non-linearity increases faster than the wave height is shallow water depths. 

It is also observed in Fig.\ref{fig:eps-hvar} that when varying the wave height, the value of the maximal steepness varies between 0.43 and 0.6, whatever the depth $h'$. The wave steepness is thus a good indicator of the non-linearities, even if the maximal slope is the most relevant one, as noticed in Fig. \ref{fig:lambda-hvar}.

\subsubsection{Choice of $N_1$ and $N_2$}
\label{N1_N2}

%\textbf{$+$ $ka=f(max\left|\frac{\partial \eta}{\partial t}\right|)$ ??}\\

As previously,  the  whole range of relative water depths was covered from shallow to deep water, as presented in Tab.~\ref{tab:kh}.
In this section, the following numerical parameters have been used:
\begin{itemize}
	\item $\epsilon^{rel}_F  = 10^{-12}$, (option: eps\_err) 
	\item $\max(\epsilon_F)  = 10.0$ (option: err\_max) 
	\item  $\epsilon^{rel}_Z  = 7 10^{-14}$  (option: eps\_inc)
	\item $\epsilon_{N_1}   =  10^{-14}$ (option: eps\_N1)
%	\item option\%itermax  = 1000.
\end{itemize}

For a given wave period $T$, a given wave height $H$ and a given water depth $h$, one can evaluate the non-dimensional numbers $H'$ and $h'$. Then, thanks to Fig.~\ref{fig:eps-hvar}, one can deduce the maximal slope. 

%As presented in section \ref{improvements}, the value of $N_1$ is adapted automatically so that the amplitude of the last mode reaches the accuracy given in input (option\%eps\_N1). 

%In order to achieve the convergence on the criteria option\%eps\_N1, the number of modes $N_1$ is given as function of  $H'$ and $h'$ (or $kH$ and $kh$ respectively) in Fig.\ref{}. \\

\begin{figure}[!htbp]
	\centering
		\includegraphics[width=\textwidth]{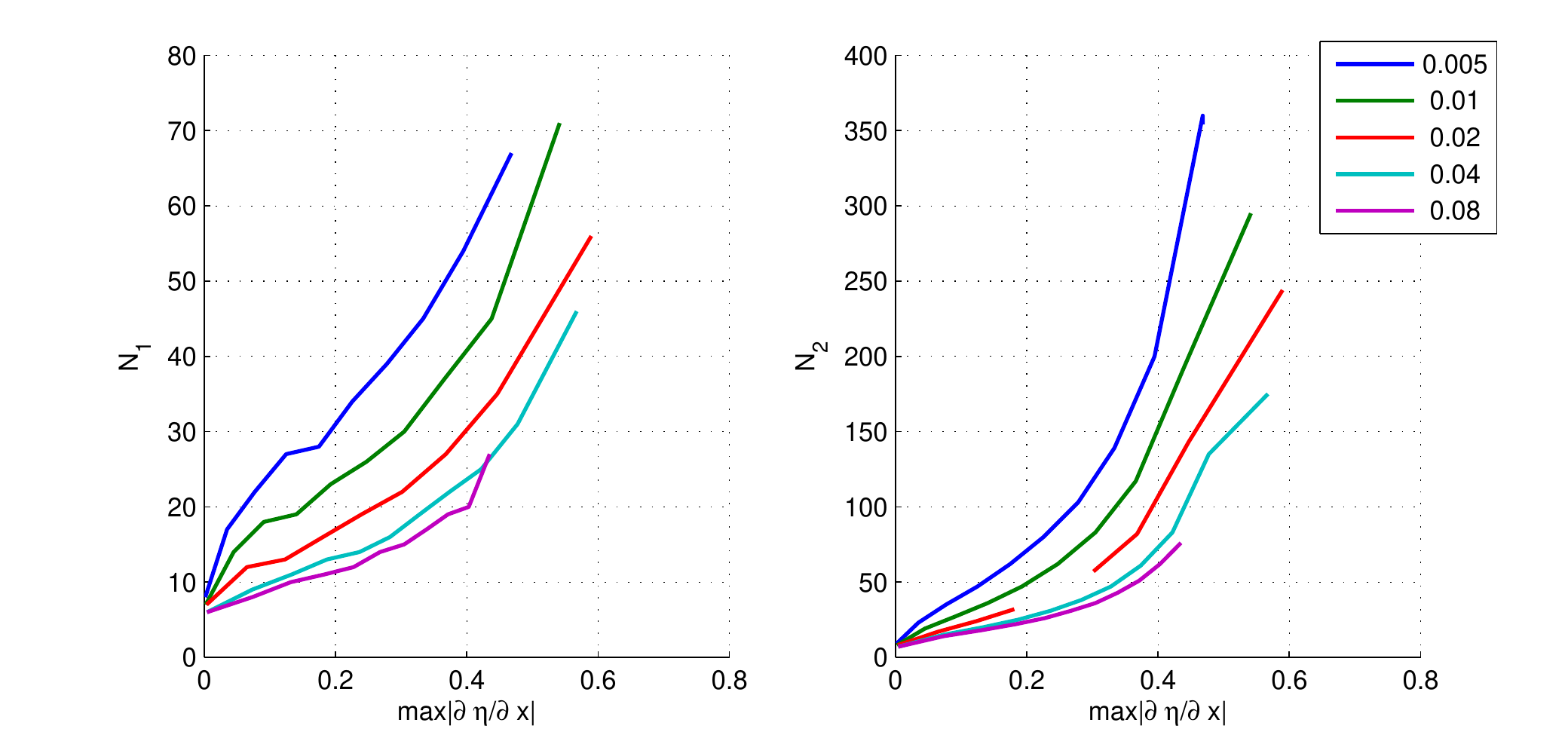}
	\caption{Necessary modes for the velocity potential ($N_1$, left) and for the elevation ($N_2$, right) for a wave propagating over a finite depth (the legend gives the values of $h'=h/gT^2$).}
	\label{fig:N1-N2}
\end{figure}

In order to achieve the convergence on the amplitude of the modes (input parameter option: eps\_N1), the number of modes $N_1$ and $N_2$ are plotted as a function of the slope in  Fig.\ref{fig:N1-N2}. It can be observed, as expected, that when increasing the slope, an increased number of modes is necessary. This is associated to the need of a larger number of modes in shallow water depth than in infinite depth at a given slope. For instance, for the maximal slope achievable, 50 (200) modes for  $\phi$ ($\eta$)  are necessary in shallow water depth and  20 (60) in infinite depth.

\begin{figure}[!htbp]
	\centering
		\includegraphics[width=0.5\textwidth]{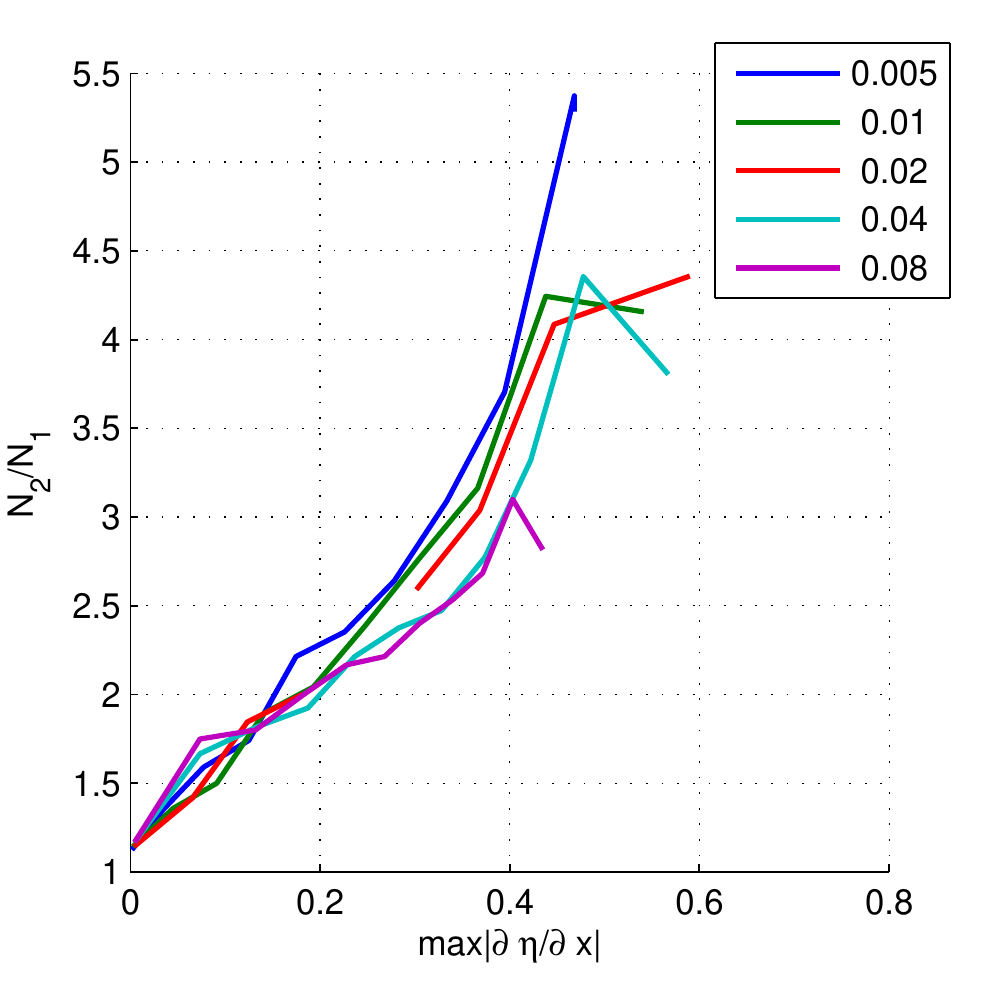}
	\caption{Ratio of $N_2/N_1$ for a wave propagating over a finite depth (the legend gives the values of $h'$).}
	\label{fig:N1oN2}
\end{figure}

As a matter of simplification of the numerical procedure, Fig.~\ref{fig:N1oN2} shows the evolution of the ratio $N_2/N_1$ as a function of the slope. 

We observe that this ratio $N_2/N_1$ is almost constant for any water depth. This allows us to extract the following relationship between those two number of modes:
\begin{equation}
N_2 \egal N_1 \left(1.5 + \frac{1.5}{0.3}\max \left|\frac{\partial \eta}{\partial x}\right|\right).
\end{equation}

This reduces to only one parameter the procedure for an automatic choice of the number of modes, as described in \ref{subsec:automatic}.

\subsection{Kinematics and pressure inside the domain}
This final section presents some examples of velocity and pressure fields obtained with CN-Stream. This illustrates the possibilities of the numerical model to provide informations about the incident wave field in view of possible coupling with CFD software for wave-structure interactions modeling.

%\textbf{add some figures of velocity and pressure fields obtained for different depths in the ka ranges presented before.}

\subsubsection{Finite water depth}

\begin{figure}[!htbp]
	\centering
	\includegraphics[width=\textwidth]{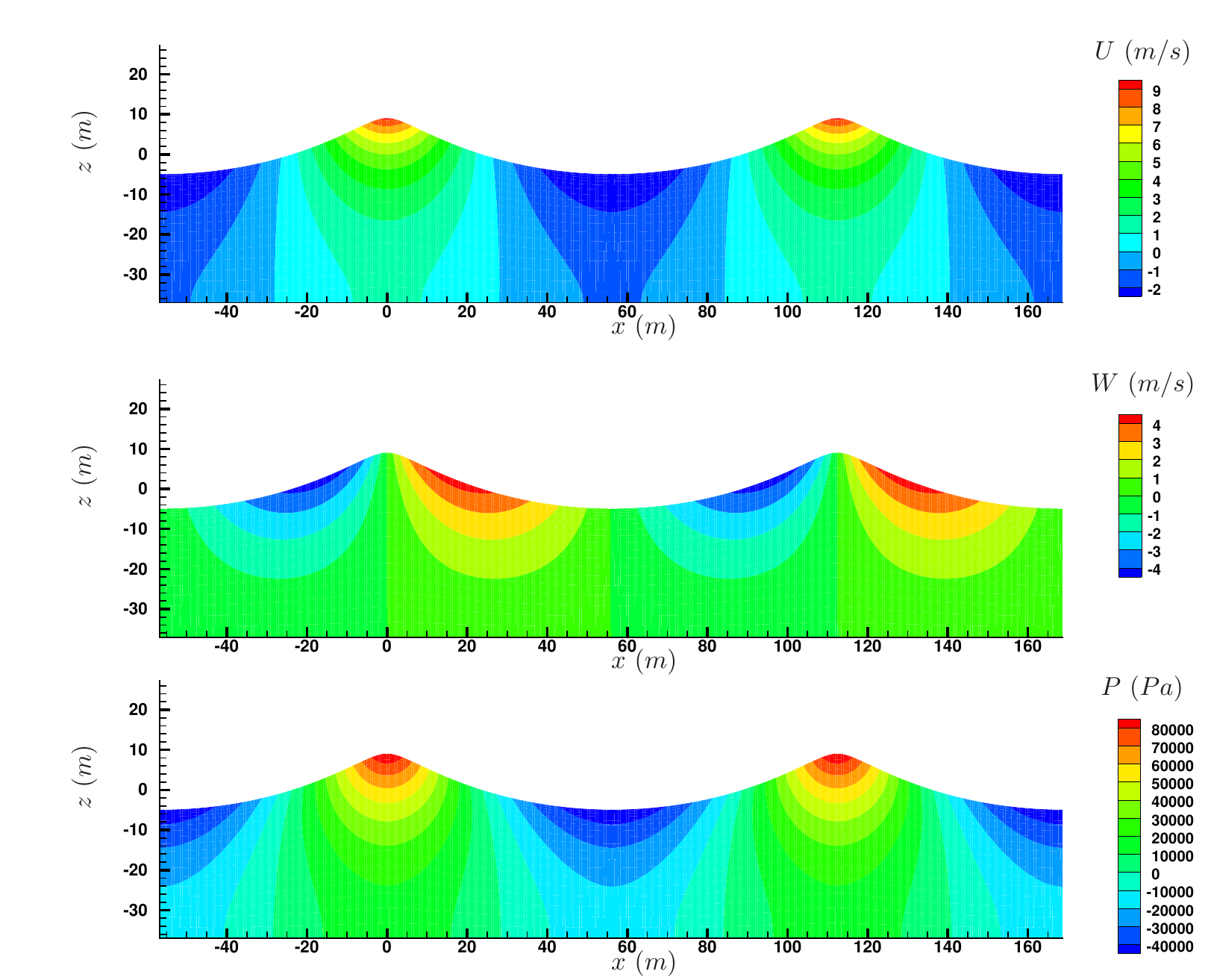}
%		\psfragfig*[width=\textwidth]{Figures/Velocities_pressure_kh2p0}{
%		\psfrag{a}[][]{$U~(m/s)$}
%		\psfrag{b}[][]{$W~(m/s)$}
%		\psfrag{c}[][]{$P~(Pa)$}	
%		\psfrag{x}[l][]{$x~(m)$}
%		\psfrag{z}[l][]{$z~(m)$}
%		}
	\caption{Horizontal velocity field (up), vertical velocity field (middle) and dynamic pressure field (bottom) for a wave propagating over a finite water depth $kh=2$ and $kH=0.78$.}
	\label{fig:volumic_field_finite_depth}
\end{figure}

The finite water depth case presented in Tab.~\ref{tab:exemple} along with the option parameters used in Section \ref{N1_N2} is computed and a reconstruction of the volume fields is performed, as presented in Fig.~\ref{fig:volumic_field_finite_depth}.

%\begin{figure}[!htbp]
%	\centering
%		\includegraphics[width=\textwidth]{Figures/U_finite_depth.eps}
%		\includegraphics[width=\textwidth]{Figures/V_finite_depth.eps}
%		\includegraphics[width=\textwidth]{Figures/P_finite_depth.eps}
%	\caption{Horizontal velocity field (up), vertical velocity field (middle) and dynamic pressure field (bottom) for a wave propagating over a finite water depth $kh=2$ and $kH=0.78$.}
%	\label{fig:volumic_field_finite_depth}
%\end{figure}

The horizontal velocity appears highly non-linear with large differences between the values in the crests and in the troughs ($\max (U) \simeq 9~m/s$ and $\min (U) \simeq -3 ~ m/s$). Similarly, the dynamic pressure field exhibits larger absolute values in the crests than in the troughs (difference is around $80~\%$).

\subsubsection{Shallow water depth}

\begin{figure}[!htbp]
	\centering
	\includegraphics[width=\textwidth]{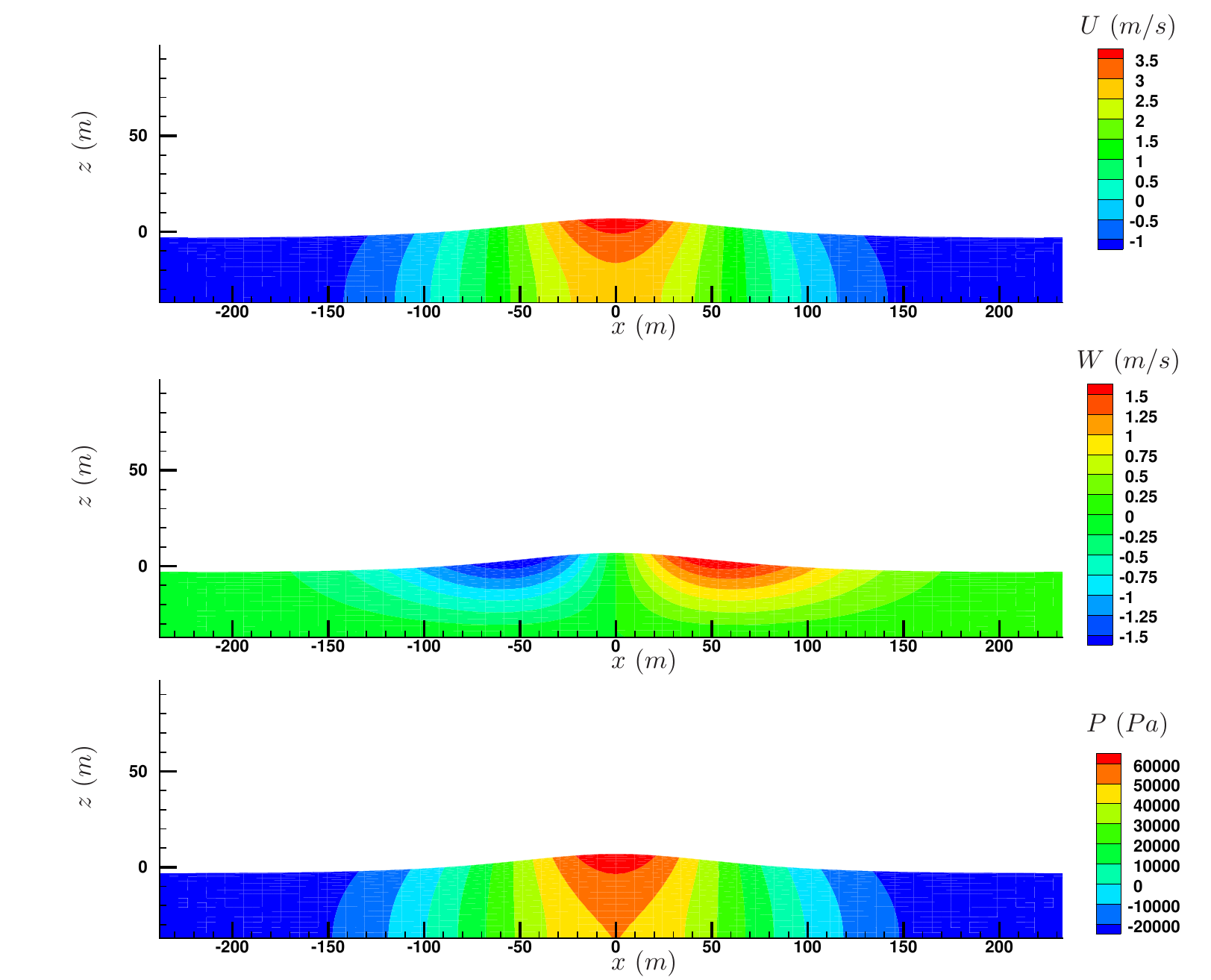}
%		\psfragfig*[width=\textwidth]{Figures/Velocities_pressure_kh0p48}{
%		\psfrag{a}[][]{$U~(m/s)$}
%		\psfrag{b}[][]{$W~(m/s)$}
%		\psfrag{c}[][]{$P~(Pa)$}	
%		\psfrag{x}[l][]{$x~(m)$}
%		\psfrag{z}[l][]{$z~(m)$}
%		}
	\caption{Horizontal velocity field (up), vertical velocity field (middle) and dynamic pressure field (bottom) for a wave propagating over a shallow water depth $kh=0.48$ and $kH=0.13$.}
	\label{fig:volumic_field_shallow_depth}
\end{figure}

The shallow water depth case presented in Tab.\ref{tab:exemple} along with the option parameters used in Section \ref{N1_N2} is computed and a reconstruction of the volumic fields is performed, as presented in Fig.~\ref{fig:volumic_field_shallow_depth}.

The non-linear features observed previously at a larger relative water depth are further enhanced with the reduced water depth. The strong asymmetry in the free surface profile (both in horizontal and vertical directions) is also observed in both the velocity and the pressure field.\\

The necessary use of fully non-linear potential wave theories in the context of highly non-linear waves, close to the wave breaking limit, is clearly demonstrated. The wave kinematics and induced pressure fields are strongly influenced by the wave non-linearity.

%\begin{figure}[!htbp]
%	\centering
%		\includegraphics[width=\textwidth]{Figures/U_shallow_depth.eps}
%		\includegraphics[width=\textwidth]{Figures/V_shallow_depth.eps}
%		\includegraphics[width=\textwidth]{Figures/P_shallow_depth.eps}
%	\caption{Horizontal velocity field (up), vertical velocity field (middle) and dynamic pressure field (bottom) for a wave propagating over a shallow water depth $kh=0.48$ and $kH=0.13$.}
%	\label{fig:volumic_field_shallow_depth}
%\end{figure}
%
%
%
%
%
%
%\clearpage
\section{Program documentation}
\label{programDoc}
%\textbf{Following paragraph not clear}\\

CN-Stream is a computational program written in Fortran language. It can be compiled as an executable file for the study of specific wave problems with inputs and dedicated outputs to be detailed in the following sections. It can also be used as a static library, which can easily be linked to other numerical models in the objective of, for instance, solve the problem of wave-structure interactions.\\

%The CN-Stream code is executed through the use of two distinct subroutines: namely $\textit{calcRF}$ and $\textit{reconstructRF}$. 
%
%$\textit{calcRF}$ is used to compute the modal amplitudes of the elevation $\eta$ and of the velocity potential $\phi$ (or equivalently the stream function $\psi$) for some given conditions. This is the core of the stream function procedure described previously.
%
%$\textit{reconstructRF}$ uses the modal amplitudes computed previously to evaluate, at a given location and at a given time, the volume fields (velocity and pressure $(U,V,W,P)$) together with the free surface elevation $\eta$ (and their derivatives if needed).
%
%Figures \ref{fig:schematic_view_calcRF} and \ref{fig:schematic_view_reconstructRF} present a schematic view of the general structure of those two main subroutines $\textit{calcRF}$ and $\textit{reconstructRF}$. More details of the different source-files are provided in next sections.
%%
%\begin{figure}[!htbp]
%\begin{center}
%\includegraphics[width=\linewidth]{Figures/schematic_view.pdf}
%\caption{Schematic view of calcRF.}\label{fig:schematic_view_calcRF}
%\end{center}
%\end{figure}
%
%\begin{figure}[!htbp]
%\begin{center}
%\includegraphics[width=0.8\linewidth]{Figures/schematic_view_reconstruct.png}
%\caption{Schematic view of reconstructRF.}\label{fig:schematic_view_reconstructRF}
%\end{center}
%\end{figure}

\begin{figure}[!htbp]
\begin{center}
\includegraphics[width=\linewidth]{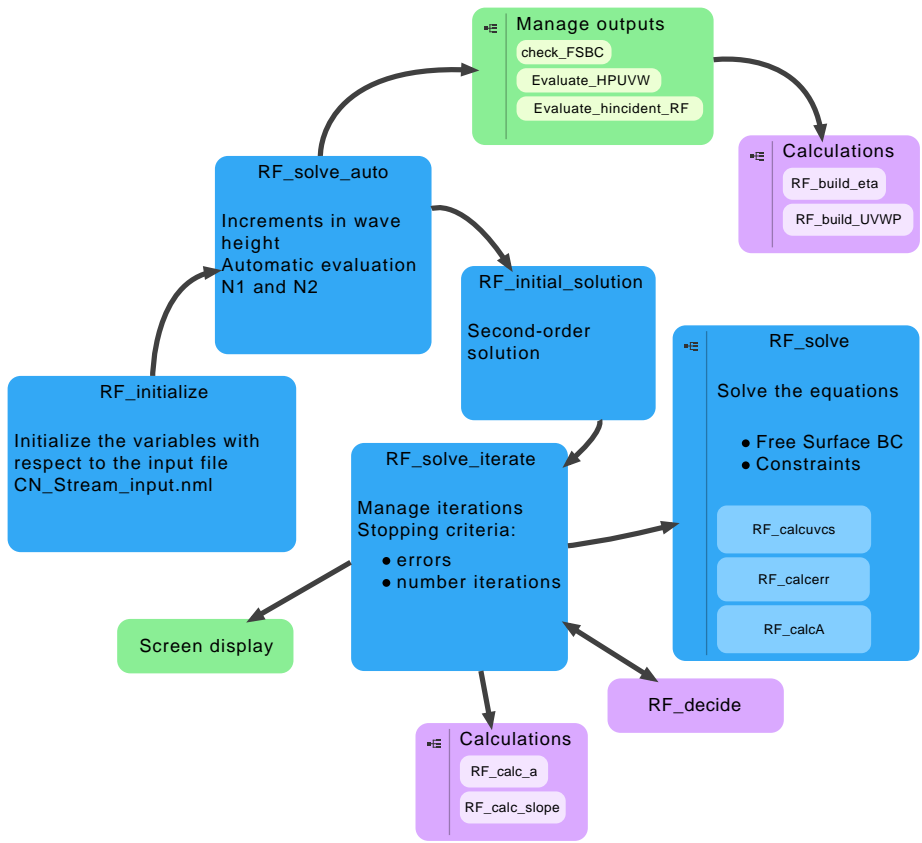}
\caption{Schematic view of CN-Stream algorithm.}\label{fig:schematic_view}
\end{center}
\end{figure}

Figure \ref{fig:schematic_view} presents an overview of the algorithm at use in CN-Stream. The main subroutines of the program are detailed with their respective purposes.

$\textit{RF\_solve\_auto}$ manages the automatic calculations of the numerical parameters at use in CN-Stream. This contains the specific enhancements proposed in the code as detailed in Sec. \ref{improvements}.

$\textit{RF\_solve\_iterate}$ manages the iterations in the solution procedure and the corresponding stopping criteria relative to the errors/tolerances (minimum amplitude of the modes, inversion of the system, etc.) as well as  the maximum number of iterations.

$\textit{RF\_solve}$ is the effective solution of the linear system of equations described previously: it is the core of the original stream function procedure.
\\

Note that for internal communications and library use in other Fortran programs, CN-Stream uses Fortran types to reduce the number of passing arguments. An example is also provided to link the library with C++ program (in particular OpenFOAM), in this case the library is interrogated to provide flow quantities at a certain position and time.
%The data communication between the main functions of CN-Stream and between CN-Stream used as a library and an external program is performed using fortran types to reduce the number of variables exchanged. 

\subsection{Source files}

\subsubsection{Project organisation and dependency}

The main folder consists in:  
\begin{itemize}
\item \texttt{CMakeLists.txt}
\item \texttt{example} Folder with examples of using the library through the communication module from Fortran and C++ 
\item \texttt{src} Folder with source files ( include also the sources of libFyMc)
\item \texttt{input} Folder with input file example
\item \texttt{output} Default folder output
\end{itemize}

The code use the library libFyMc to read the ``dictionary'' input file. The library is provided with the sources.

\subsubsection{CN-Stream  - variables and types}

The different Fortran types (RF\_type, option\_type, output\_type) are defined explicitely in \texttt{variables\_CN\_Stream.h} and \texttt{variables\_output\_CN\_Stream.h} and are included when needed in CN-Stream. It allows the user to include them easily into CN-Stream but also in another code, which makes use of the CN-Stream library.\\

\noindent
In more details, those types include:
\begin{itemize}
	\item RF\_type
	\begin{itemize}
		\item definition of the parameters of the wave, corresponding to the input parameters specified in the input file as detailed in Sec. \ref{subsubsec:Input}
		\item Modal amplitudes of the free surface elevation $\eta$ and of the velocity potential $\phi$ (or equivalently the stream function $\psi$).
		\item If needed, free surface elevation and slope in the spatial domain.
	\end{itemize}
	\item option\_type: all options relative to the solution method, specified in input file as detailed in Sec. \ref{subsubsec:Input}. This type also includes the optimal number of points $N_1$ and $N_2$ resulting from the procedure detailed in Sec. \ref{subsec:automatic}.
	\item output\_type: defines for one location $(X,Y,Z)$ the free surface elevation, the pressure and velocity components together with the necessary time and/or spatial derivatives of those components. The possible existence of a $Y$-component is associated to the definition of an angle of propagation as input, referenced as $\theta$.
\end{itemize}

\subsubsection{CN-Stream - main program}

The set of Fortran files needed in order to compile CN-Stream is listed in Tab. \ref{tab:source_files} with a brief description of the purpose of each of the source file.\\
\begin{table}[h!tbp]
\begin{tabular}{l p{0.6\textwidth}}
%\hline 
%\multicolumn{2}{|l|}{CN\_Stream source files}  \\ 
\hline 
%\multicolumn{2}{|l|}{CN\_Stream source files}  \\ 
{CN-Stream source files} & \\
\hline
\textbf{mod\_CN\_Stream.f90} & Module allowing communication without using complex datatypes (C++) \\
%\hline 
\textbf{main\_CN\_Stream.f90} & Main program for CN-Stream computations  \\ 
%\hline 
\textbf{modSolve.f90} & Solves the equation of the problem described above \\ 
%\hline 
\textbf{modCNinitialize.f90} & Initialization of CN-Stream computation \\ 
%\hline 
\textbf{modUtils.f90} & Useful functions \\ 
%\hline 
\textbf{modMatrix.f90} & Computes the inverse matrix from the least square method \\ 
%\hline 
\textbf{modType.f90} & Definition of types and useful constants \\ 
%\hline 
\textbf{modModal.f90} & Useful functions used in modSolve.f90\\ 
%\hline 
\textbf{HOS\_modlinear\_wave.f90} & Computation of linear dispersion relation \\ 
%\hline 
\textbf{HOS\_modmaths.f90} & Useful mathematical functions \\ 
%\hline 
\textbf{modSetupNameList.f90} & Read the input NML file \\ 
%\hline 
\textbf{modReconstrucVol.f90} & Evaluate wave elevation, pressure, velocity and its derivatives  \\ 
%\hline 
\textbf{modReconstruction.f90} & Recompute wave elevation, pressure, velocity  and wave slope from Fourier coefficients and the other way around .\\ 
%\hline 
\textbf{modOutputs.f90} &  Write outputs on files.\\ 
\textbf{variables\_CN\_Stream.h} & Definition of Fortran types: RF\_type and option\_type \\
\textbf{variables\_output\_CN\_Stream.h} & Definition of Fortran type: output\_type \\
\hline 
\end{tabular}
\caption{List of source files used in CN-Stream}\label{tab:source_files} 
\end{table}

\subsubsection{CN-Stream - library}

For the possible use of CN-Stream as a library in another program, the set of Fortran files is similar to the one described in previous subsection.  There are two ways to use CN-Stream as a library.

\begin{itemize}
\item Use the declarations of the variables of the \texttt{variables\_CN\_Stream.h} and \texttt{variables\_output\_CN\_Stream.h} and call the functions described in the source file \texttt{lib\_CN\_Stream.f90}.
\item Use the subroutines indicated in the communication module \texttt{mod\_CN\_Stream.f90}. 
\end{itemize}

%  $calcRF(ConfigFile, RF, option)$ 

% $reconstructRF(RF, option, x, y, z, time, thetaincident, hydrostatic, output)$ 
%Ces routines ont comme param�tres d'entr�es et de sorties:
%\begin{itemize}
%   \item ConfigFile (input) : le fichier $RF\_input.nml$,
%   \item RF (output pour $calcRF$ et inout pour $reconstructRF$) : de type $RF\_type$,
%   \item option (output pour $calcRF$ et input pour $reconstructRF$) : de type $option\_type$,
%   \item x, y, z (input) : r�els d�finissant la position � laquelle on veut reconstruire les grandeurs volumiques,
%   \item time (input) : r�el d�finissant  le temps auquel on veut reconstruire les grandeurs volumiques,
%   \item thetaincident (input) : r�el d�finissant l'angle d'incidence de la houle,
%   \item hydrostatic  (input) : variable logique permettant la prise en compte ou non de l'hydrostatique,
%   \item output (output) : de type $output\_type$.
%   \newline
%\end{itemize}

\subsection{Compilation}

The code can be compiled on any computer architecture. One only needs a Fortran compiler (for instance gfortran, the GNU Fortran compiler, part of GCC).  A makefile is provided but the recommended procedure is to use cmake.
The following commands can be executed in the root folder where \texttt{CMakeLists.txt} is located,  to compile the dependency, the executable and the shared library:
\begin{itemize}
\item  \texttt{cmake -H. -Bbuild} 
\item  \texttt{cmake -{}-build build}
\end{itemize}

%{\color{red} It is necessary to do a make in the build directory after that?!! +specify that these first commands need to be done at the root of the folder +detail where is the mainCNS created?
%}
%\item \underline{Library:} \texttt{make createlib\_Debug} or \texttt{make createlib\_Release}

%A \texttt{Makefile} is provided with the source code for compiling CN-stream (as execuatble or library) under Linux/Unix environment. The following commands can be used:
%\begin{itemize}
%\item \underline{Executable:} \texttt{make} command executed at the root of the project will create the binary \texttt{CN\_Stream},
%\item \underline{Library:} \texttt{make createlib\_Debug} or \texttt{make createlib\_Release} command executed at the root of the project will create the static library file \texttt{lib\_CN\_Stream.a}. This static library can be used in the compilation stage of external programs that intends to embed CN\_Stream. The only compilation options differ between the two commands provided.
%\end{itemize}

Compilation has been tested with gfortran on different Unix/Linux platforms as well as in Windows environment.\\

For Windows environment, compilation using Intel Visual Studio has also been tested. The program is provided with the corresponding project file \texttt{CN\_Stream.vfproj} allowing a straightforward compilation of the code.

\subsection{Running CN-Stream}
\label{subsec:Running_CN}

CN-Stream has been developed for command-line run with an input file located in the \texttt{input} folder containing all specifications needed. All output files will be created in the directory \texttt{output}, but other specifications can be given. Details of inputs and outputs are provided hereafter. The executable can be run with the command \texttt{./mainCNS}. The name of the dictionary can be specified as an argument. 

\subsubsection{Inputs}
CN-Stream needs as input the characteristics of the wave, together with some informations relative to the numerical solution of the problem (target accuracy, etc.). The wave can be described in dimensional or non-dimensional form. As a matter of clarity, the wave parameters to provide as input are detailed in the next paragraphs depending on the need of the user. Note that those parameters are provided within an input file which content is also detailed.\\

In CN-Stream, the non-linear regular water wave is characterized by:
\begin{itemize}
	\item the water depth $h$, possibly infinite,
    \item the wave length $\lambda$ \textbf{or} the wave period $T$,
	\item the wave height $H$ (distance from crest to trough),
	\item the constant current superimposed to the wave (under the form of a Eulerian current or a given transport of mass).
\end{itemize}

\paragraph{Input file}
\label{subsubsec:Input}

The input file is assumed to be named \texttt{CN\_Stream\_input.dict}. Table \ref{tab:input_file} describes the different parameters accessible in this input file. The Options\_solver parameters are useful for an advanced user, in order to obtain solutions with a controlled accuracy and/or to look for waves close to the wave breaking limit.

\begin{table}[h!tbp]
\begin{tabular}{l p{10cm}}
\hline
\multicolumn{2}{l}{waveInput: Label (\texttt{waveStream} in example file) and Definition of the characteristics of the simulated wave}  \\ 
\hline 
\textbf{GeneralDimension} & Dimensional (=1) or Non-dimensional (=0) \\ 
%\hline
\textbf{GeneralDepth} & h if GeneralDimension=1 /
                              h' if GeneralDimension=0 \\ 
%\hline 
\textbf{GeneralModes} &  Number of modes for first evaluation\\ 
%\hline
\textbf{WaveInput} & Period  / Wavelength  (if GeneralDimension=0  only Wavelengthis  possible) \\ 
%\hline
\textbf{Period} &  period value if WaveInput set to Period\\ 
%\hline
\textbf{Wavelength} &  Wavelength value if WaveInput set accordingly \\ 
%\hline
\textbf{WaveHeight} & H if GeneralDimension=1 / H'
                           if GeneralDimension=0\\ 
%\hline
\textbf{CurrentValue} & value of current ; dimensional if input: GeneralDimension=1
/ non-dimensional if input: GeneralDimension=1\\ 
%\hline
\textbf{CurrentType} & type of current ; 1 mass transport / 0 eulerian current \\ 
\hline
\multicolumn{2}{l}{Options for the numerical solution of the problem}  \\ 
\hline
\textbf{n\_H} & $n_H$: Number of steps in wave height \\ 
%\hline
\textbf{err\_type} & Error type: 0 absolute ; 1 relative\\ 
%\hline
\textbf{eps\_err} & $\epsilon^{rel/abs}_F$ Tolerance on the equations \\ 
%\hline
\textbf{err\_max} &  $\max(\epsilon_F)$: Error value over which computation is considered divergent\\ 
%\hline
\textbf{eps\_inc} & $\epsilon^{rel/abs}_Z$: Convergence criteria on the unknowns\\ 
%\hline
\textbf{eps\_N1}  & $\epsilon_{N1}$: Decision criteria on the modes for the automatic adjustment of $N_1$\\
%\hline
\textbf{itermax}  & Maximum number of iterations\\ 
%\hline
\textbf{increment\_type} & choose between a linear or exponential incrementation:  Increment type for wave height / 0 linear ; 1 exponential\\ 
%\hline
\textbf{printonscreen} &print the intermediate results of the simulation on the command prompt: Print on screen =1 / do not print on screen = 0\\ 
%\hline
\textbf{writeoutput} &  Write output files =1 / do not Write output files = 0\\ 
%\hline
\textbf{subdict} &  Standard way to include dictionnary (here ``Output'' in another using libFyMc) \\ 
\hline
\multicolumn{2}{l}{Outputs: supplementary info for outputs }  \\ 
\hline
\textbf{Path} &  Specify output path (default: ``./output/'')\\ 
%\hline
\textbf{x/y/z/time} &  Specify position to evaluate quantities for local outputs (default:``0.0'') \\ 
%\hline
\textbf{theta} &  Incident angle of waves (default:``0.0'') \\
\hline
\end{tabular}
\caption{Description of input file parameters for CN-Stream}\label{tab:input_file} 
\end{table}

\subsubsection{Output files}

Depending on the choices made in the input file (see Tab.\ref{tab:input_file}), different output files are created. They are located at the root of the folder. Input file also defines if outputs are dimensional or non-dimensional quantities. Following files may be created:
\begin{itemize}
	\item \texttt{waverf.cof} gives the main important parameters of the simulation, namely $\lambda$, $H$, $k$, $T$, $c$, $c_S$, $c_E$, $N_1+1$, $N_2+1$, $R$, $h$ (in dimensional or non-dimensional form depending on the value of input: GeneralDimension in the input file) as well as the modal amplitudes $a_n$ and $b_n$,
	\item \texttt{waverf.dat} gives the modal amplitudes $a_n$ and $b_n$.
\end{itemize}

In complement, different subroutines may be called to write the necessary outputs needed by the user. They are available inside the source files and a simple call in the main program will enable the corresponding outputs:
\begin{itemize}
	\item \texttt{WriteOutput}: this subroutine creates the file \texttt{resultsOutput.txt} containing at a given location and time all spatial quantities computed by CN-Stream (free surface elevation, velocities, pressures, derivaitves, etc.).
	\item \texttt{TecplotOutput\_Modes}: this subroutine creates the file \texttt{Modes\_CN\_Stream.dat} containing the modal description of the free surface elevation and velocity potential, for use with Tecplot.
	\item \texttt{TecplotOutput\_VelocityPressure}: this subroutine creates the file \texttt{VP\_card\_fitted.dat} containing the velocity and pressure field under the simulated wave, for use with Tecplot.
	\item \texttt{TecplotOutput\_FreeSurface}: this subroutine creates  the file \texttt{FreeSurface\_CN\_Stream.dat}, which provides the free surface elevation and slope.
\end{itemize}

\section{Conclusions}

CN-Stream has been developed to compute non-linear regular ocean waves with a high level of accuracy. The model is limited to arbitrary constant water depth and non-breaking waves. CN-Stream is an open-source code, redistributed under the terms of the GNU GPL v3 License as published by the Free Software Foundation. It is available through the GitHub platform \cite{CN-Stream}. Along with the source code, a Wiki documentation is available, which makes the compilation  of the source files and the execution easy. It can be used as a stand-alone binary or as a library to be included in another program.\\

The code is based on the stream function theory and the original works of \cite{Rienecker} and \cite{fenton1988numerical} have been taken as basis. Some enhancements are proposed in the current implementation, namely: i) a possible different number of modes to represent the free surface elevation and the stream function (or eq. the velocity potential) and ii) an automatic calculation of the optimal number of collocation points (or equivalently of the number of modes) to reach a target accuracy.

It has been demonstrated that these allow an increase accuracy of the numerical solution, together with an extended domain of application with respect to the maximum wave height accessible.\\

Different example of applications of the model are provided, which demonstrate the importance of the non-linear effects in the description of a regular waves. The free surface profiles are analyzed over a wide range of steepness and relative water depth, together with the kinematics and pressure fields.

The volume fields are a standard output of CN-Stream, which intends to provide a simple and accurate code for the description of non-linear regular waves, especially in the context of wave-structure interactions. It is worth noticing that it is already at use in different codes dealing with wave-structure interactions at Centrale Nantes. Among other, CFD codes such as ICARE \cite{reliquet_simulation_2013, reliquet_assessment_2017}, WCCH \cite{li_VOF_2017} or OpenFoam \cite{li_progress_2017} uses CN-Stream as a library for the description of incident non-linear regular waves.

%HOS-ocean code has been developed to study the propagation of large-scale nonlinear sea states. HOS-ocean is an open-source code, redistributed under the terms of the GNU General Public License (version 3) as published by the Free Software Foundation. Along with the source code, a documentation under wiki format is available which makes compilation and execution of the source files easy. In addition, benchmarking procedures as well as typical post-processing examples are provided with the main code.

%The code has been shown to be accurate and highly efficient thanks to the interesting convergence properties of pseudo-spectral methods. It has been highly validated on several configurations and one of the purpose of the open-source release is to encourage researchers to use it. Particularly, thanks to the post-processing provided, the coupling between HOS-ocean and CFD software for wave-structure interactions modeling should be straightforward.

\section*{References}
%%%%%%%%%%%%%%%%%%%%%%%%%%%%%%%%%%%%%%%%%%%%%%%%%%%%%%%%%%%%%
%% BIBLIOGRAPHY AND OTHER LISTS
%%%%%%%%%%%%%%%%%%%%%%%%%%%%%%%%%%%%%%%%%%%%%%%%%%%%%%%%%%%%%
%% A small distance to the other stuff in the table of contents (toc)
%\addtocontents{toc}{\protect\vspace*{\baselineskip}}

% The Bibliography
% ==> You need a file 'literature.bib' for this.
% ==> You need to run BibTeX for this (Project | Properties... | Uses BibTeX)
%\addcontentsline{toc}{chapter}{Bibliography} %'Bibliography' into toc
%\nocite{*} %Even non-cited BibTeX-Entries will be shown.
%\bibliographystyle{alpha} %Style of Bibliography: plain / apalike / amsalpha / ...


\bibliographystyle{plain}
\bibliography{literature} %You need a file 'literature.bib' for this.

\end{document}